\documentclass[review]{elsarticle}
\usepackage{cancel}
\usepackage{ulem}
\usepackage{soul}
\usepackage{lineno,hyperref}
\usepackage{lscape}
\usepackage{pdflscape}
\usepackage{subcaption}
\usepackage{adjustbox}
\usepackage{url}
\usepackage{amssymb,amsmath, amsthm, setspace,graphicx}
\usepackage{graphicx}
\usepackage{float}
\usepackage{placeins}
\usepackage{multirow}
\usepackage[labelfont={bf,small}]{caption}
\usepackage[font={small}]{caption}
\usepackage[toc,page]{appendix}
\usepackage{setspace}
\usepackage[T1]{fontenc}
\usepackage[margin=1in]{geometry}
\usepackage{comment}
\begin{document}

\begin{centering}
\title{ \bf Equity Premium Prediction: \\Taking into Account the Role of Long, even Asymmetric, Swings\\ in Stock Market Behavior}

\vskip2.5cm
\author{  Kuok Sin Un $^{a}$  and Marcel Ausloos $^{a,b,c} $ 
 \\
 \vskip1.5cm
 $^a$
School of Business, University of Leicester,\\
Brookfield,  Leicester, LE2 1RQ, UK\\   %e-mail: ma683@leicester.ac.uk\\
 $^b$ 
 Department of Statistics and Econometrics,  \\ Bucharest University of Economic Studies,  15-17 Dorobanti Avenue, \\ District 1, 010552, Bucharest, Romania, \\  %e-mail: marcel.ausloos@ase.ro \\
  $^c$
Group of Researchers Applying Physics in Economy and Sociology (GRAPES),  \\Beauvallon, 
Sart Tilman, B-4031, Li\`ege Angleur, Belgium, Europe  
 % e-mail: marcel.ausloos@uliege.be
 \vskip1.5cm
%$^{*}$Corresponding author  : School of Business, University of Leicester, Brookfield,	Leicester LE2 1RQ, United Kingdom.   \\ email address: ksu1@leicester.ac.uk
}
\date{\today}
\end{centering}

%\newpage
\begin{abstract}
 Through a novel approach,  this paper shows that substantial change in stock market behavior has a statistically and economically significant impact on equity risk premium predictability both on in-sample and out-of-sample cases. In line with Auer's ``B ratio'', a ``Bullish index'' is introduced to measure the changes in stock market behavior, which we describe through a "fluctuation detrending  moving average analysis" (FDMAA) for  returns.  We consider 28 indicators.  We find that a ``positive shock'' of the Bullish Index is closely related to strong equity risk premium predictability for forecasts based on macroeconomic variables for up to six months. In contrast, a ``negative shock'' is associated with strong equity risk premium predictability with adequate forecasts for up to nine months when  based on technical indicators.
 
\end{abstract}
\maketitle

%\begin{keyword}
%Hurst exponent, Fluctuation Detrending Moving Average Analysis, Bullish Index, Equity Risk Premium Predictability, Stock Market Behavior
%\end{keyword}

%\linenumbers
\newpage
\section{Introduction}\label{sec:Introduction}

\pagenumbering{arabic}
The long-range dependence correlations in particular in time series of financial assets have received  much attention through fractal (and multifractal) aspects \vphantom{\citep{ivanova1999low,ausloos2000statistical,ivanova2002eur,calvet2002multifractality,BuonocoreBrandiMantengaDiMatteo2020,KukackaKristoufek2020,BrandiDiMatteo2021,chan2021economic}}  
\citep{ivanova1999low}-\citep{chan2021economic}
The so called strength of long-range dependence is traditionally measured by the Hurst exponent ($H$) \citep{hurst1951long}. Along such a line, extensive research has been done on the scaling behavior of various asset classes \citep{jiang2019multifractal} such as precious metals \citep{wu2017fractal}, currency exchange rates \vphantom{\citep{vandewalle1997coherent,ausloos2004classical, cristescu2012parameter}}
\citep{vandewalle1997coherent}-\citep{cristescu2012parameter}, stock market indices \vphantom{\citep{ausloos2002multifractal,alvarez2008time,wang2009analysis,zhu2018multifractal}}
\citep{ausloos2002multifractal}-\citep{zhu2018multifractal}, crude oil prices \vphantom{\citep{yang2016multifractal,watorek2019multifractal,yao2020multifractal}}
\citep{yang2016multifractal}-\citep{yao2020multifractal} or cryptocurrencies \citep{takaishi2018statistical,zhangstylised}, among many other concerns.

Recent existing studies tend to focus on the use of the Hurst exponent in relation to time-series forecasting models.  We notice that the use of macroeconomic variables, including interest rates, valuation ratios, and bond spreads have received significantly less attention in the studies based on the Hurst exponent. The objective of the present paper is to  report results of a study investigating the impact of substantial changes in stock market behavior \footnote{We define a stock market behavior in relation to the level of the Hurst ($H$) exponent itself. Let it be recalled that the Hurst exponent lies in the range $0<H<1$: for  an "antipersistent behavior", the value of the Hurst exponent is $0<H<0.5$, which implies that an increment (decrement)  has a higher than 50\% probability to be followed by a decrement (increment). {\it A contrario}, the value of the Hurst exponent  in the range $0.5<H<1$ denotes a "persistent behavior", i.e., an increment (decrement) has a higher than 50\% probability to be followed by an increment (decrement); a Brownian motion is characterized by $H=0.5$. } on equity risk premium predictability with forecasts based either on macroeconomic variables or on technical indicators, specifically, but  without much limitation, in the U.S. market. 

The method goes as follows. First, we calculate the daily local Hurst exponent using the  ``fluctuation detrending moving average analysis'' (FDMAA) method,  e.g. see  Ref. \vphantom{\citep{vandewalle1998crossing,di2005long,carbone2009detrending,gu2010detrending,chen2016finite}} \citep{vandewalle1998crossing}-\citep{chen2016finite}.

Then, we construct a Bullish ratio, motivated by Auer \citep{auer2016performance}:  it is defined as the ratio between the number of positive returns (and when $H>0.5$) to the total number of returns in a given period. In this paper, for space saving, we focus on positive returns and $H>0.5$, - mainly in order to emphasize a market index with a high  Hurst exponent  level, whence  associated with strong return predictability \citep{eom2008hurst}; the other variants can be easily worked out within the proposed method, but are left  for further applications by interested readers. A Bullish Index is constructed next to measure the changes in the Bullish ratio; details are found in section \ref{subsec:Construct the Bull Index}. We distinguish in-sample  and out-of-sample cases   when  investigating the spikes of the Bullish Index. 
It seems obvious that a substantial switch of market behavior,  a posteriori indicated by a spike,  should have  important implications for investors as it tends to be associated with financial market instability \vphantom{\citep{grech2004can,grech2008local,varela2015long}}\citep{grech2004can}-\citep{varela2015long}. 

Thereby, it is shown below that we make two major contributions to the existing empirical literature, beside providing a theoretical method. First, we find that a sharp rise or deep fall of the Bullish Index captures major changes in stock market behavior, - which   often  tend to occur before or at   economic recessions. We call the peak (top 2.5\% quantile) of the Bullish Index as a ``positive shock''. In contrast, the trough (bottom 2.5\% quantile) of the Bullish Index can be interpreted as a ``negative shock''. 

%{\bf I disagree; in my opinion: the Bullish index considers only positive returns;  a peak occurs when   some $B_{t-1} < B_t  $ and  when $B_t > B_{t+1}$. The persistence is demanded by the $B_t$ definition. However, if one has an antipersistent behavior (H$<$0.5), $B_t$ IS NOT DEFINED . One may generalize $B_t$  when not including the constraint on $H$; one can then have a peak in $B_t$  but  that is again due to the ratio between positive and negative returns over 3 consecutive months;  there must be a turnover, I agree. {\color{black}  \bf Agree, I have removed the lines.}

%{\it  thus I would cancel or adapt}  \cancel{[ The peaks} of the Bullish Index occur when the market switches from negative persistent behavior (where the number of negative returns with $H>0.5$ dominates in a given month) or antipersistent behavior (many trading days with $H<0.5$ in a month) to positive persistent behavior (where the number of positive returns with $H>0.5$ dominates \cancel{in a month}). ]

Secondly, we test the performance of forecasting models based  either on macroeconomic variables or on  technical indicators, in comparison to a historical average forecast as suggested by Welch and Goyal \citep{welch2008comprehensive}. The performance is measured by the "certainty equivalent return", a measure which can be interpreted as the risk-free return that a risk-averse investor is willing to accept instead of holding a risky portfolio, see details in Campbell and Thompson \citep{campbell2008predicting}. We show below that forecasts based on macroeconomic variables have substantially stronger and significant equity premium predictability in outperforming the historical average forecast for up to six months following   positive shocks. 
 
In contrast, forecasts based on technical indicators have substantially stronger and significant equity premium predictability in outperforming the historical average forecast for up to nine months following the negative shocks. Thus, these findings provide important implications for investors on the usefulness of macroeconomic variables or technical indicators as predictors in relation to changes in stock market behavior, beside adding  to a developing literature that utilizes the Hurst exponent  for investment strategies \vphantom{\citep{peters1994fractal,kim2010portfolio,domino2011use,batten2013structure}}\citep{peters1994fractal}-\citep{batten2013structure}.

The remainder of the paper is structured as follows. Section \ref{sec:method} describes the FDMAA  method and the construction of the Bullish Index. Section \ref{sec:datasum} describes the data and provides some statistical characteristics. Section \ref{sec:Empirical Results} reports the empirical results. Section \ref{sec:conclusion} concludes the paper with emphasis on our outlined original contributions.

\section{Literature Review }\label{sec:literature 2}

\subsection{Literature on Estimation of the Hurst Exponent}\label{sec:literature 2b}

The Hurst exponent  \citep{hurst1951long} can be measured from the detrended fluctuation analysis (DFA) method   \citep{ossadnik1994correlation,peng1994mosaic}; the DFA method is characterized by removing local trends with polynomial regressions.  Since then, many   techniques have been introduced which focused on a more efficient Hurst exponent estimation in financial time series  
  \citep{jiang2019multifractal}, \vphantom{\citep{di2005long,carbone2009detrending,gu2010detrending,chen2016finite}}\citep{di2005long}-\citep{chen2016finite}; the pioneering work of Vandewalle and Ausloos \citep{vandewalle1998crossing} shows that cross-overs of moving averages are useful in estimating the Hurst exponent.   Alessio et al. \citep{alessio2002second} further developed the idea and proposed the DMA method (detrended moving average) which removes local trend with moving averages. 

\subsection{Application of Hurst Exponent in Financial Markets}\label{sec:literature 2c} 

The study of long-range dependence of financial time-series was pioneered by Mandelbrot \citep{mandelbrot1971analysis}.   Since then, extensive research has been  performed on the scaling behavior of various asset classes through  the Hurst exponent; relevant to long term  memory in financial time series, one can quote  among many others \citep{vandewalle1997coherent}, \vphantom{\citep{mantegna1995scaling,amaral1997scaling,vandewalle1998multi,ausloos2002financial,cerqueti2003microeconomic,baciu2014ranking,gunduz2022entropic,tilfani2022heterogeneity}}\citep{mantegna1995scaling}-\citep{tilfani2022heterogeneity}. 
  %Gunduz2022entropic
 
Without going into lengthy details, let us claim that the existing literature shows that   the Hurst exponent contains some valuable information in forecasting asset returns and market crash analysis. However, we notice that existing studies on asset return predictability did not at all focus on the distinction between forecasting models based either on macroeconomic variables  or on technical indicators. In addition, the mixed evidence from  studies of market crashes, - huge amplitude of rather short lived spikes, implies that analyzing market efficiency solely with the Hurst exponent is not sufficient.  Thus, an alternative measure of the changes in stock market behavior is needed, is proposed below, and demonstrated to be efficient in specific limits.

\section{Methods}\label{sec:method}

\subsection{Estimate the Time-dependent Hurst Exponent}\label{subsec:Estimate Local Hurst exponent}
The daily market returns ($r(d)$) are defined as the daily logarithm returns of the closing price ($p_d$) for a stock market index : $r(d)= log (p_{d}/p_{d-1})$. We filter the daily stock market returns  based on the GARCH (1,1) model which aims to eliminate the short-range dependence of the time series \citep{cajueiro2004hurst,jafari2007does}. 

The details of the fluctuation detrending  moving average analysis (FDMAA) method are summarized in the following 5  steps as outlined by Gu and Zhou \citep{gu2010detrending}  \footnote{We choose the "position parameter" to be equal to  zero,  - which	corresponds to the backward moving average; see more details  about the position parameter in Xu et al. \citep{xu2005quantifying}.}.

Step 1. Consider a    $N$-long time series of filtered daily market returns $r(d)$, where $d=1,2,\dotsc,N$. 
The sequence of cumulative sums is constructed as

\begin{equation}\label{step1}
y(d) = \sum_{i=1}^{d} r(i), \;\; d=1,2,\dotsc,N.
\end{equation}

Step 2. Calculate the moving average function $\tilde{y}$ in a moving window of size $n$,

\begin{equation}\label{step2}
\tilde{y}(d;n) = \frac{1}{n}\sum_{k=0}^{n-1} y(d-k).
\end{equation}
where  $k$  points out to a specific point in the window.

Step 3. Detrend the signal series by subtracting  the moving average function $\tilde{y}$ from $y(i)$ and obtain the residual sequence $\epsilon(i)$,

\begin{equation}\label{step3}
\epsilon(i)=y(i)-\tilde{y}(i),
\end{equation}

for each possible value of $i$ such that  $n \leq i \leq N$.  

Step 4. Divide the residual series $\epsilon(i)$ into $N_{n}$ disjoint segments with the same size $n$, where $N_{n}$ is the integer part of $N/(n-1)$. Each segment 
can be denoted by $\epsilon_{v}$ such that $\epsilon_{v}(i)=\epsilon(l+i)$ for $1 \leq i \leq n$,  where $l=(v-1)n$. 
  
The root-mean-square ($average$)  function $F_{v}(n)$ over all windows with segment size $n$ can be calculated  from

\begin{equation}\label{step4}
F_{v}^{2}(n) = \frac{1}{n}\sum_{i=1}^{n} \epsilon_{v}^{2}(i).
\end{equation}

Step 5. Varying the %values of 
segment size $n$ for all possible values, a power–law relation  between the function    $F_{v}(n) $ and the size scale $n$ is written as follows:  

%\begin{equation}\label{step6} F_{q}(n) 	\sim  n^{H(q)}  \end{equation} 
\begin{equation}\label{Hdefeq}
 F_{v}(n) \simeq n^H  
\end{equation}

where the Hurst exponent $H $ can be estimated by the linear regression $ log \, F_{v}(n) = c + H  \, log \, n + u(n) $;  $c$ is a constant and $u(n)$ is the error term. 
% In line with \citet{kristoufek2012fractal}, we set $q=2$ which is the multifractal order corresponding to the scaling of variance. 
%Thereafter, we label $H\equiv H(2)$ further in the text. 
  In the present case, we estimate the  so called  "local Hurst exponent"  \citep{ausloos2000statistical,ausloos2003strategy} using a sliding window of size $N=215$ with a step of one day \footnote{Grech and Mazur \citep{grech2004can} recommend any choice $190\leq N \leq 230$; if N is too large, $H$ loses its locality, whereas if N is too small, this leads to a substantial drop in $R^{2}$ in a linear regression. We choose $N=215$ in line with the choice of other authors, e.g. Grech and Pamula \citep{grech2008local} and Horta et al. \citep{horta2014impact}.}. Following Horta et al. \citep{horta2014impact}, we implement the FDMAA algorithm with the parameters set as follows: $n_{min} = 5$, $n_{max}=43$,  with $\phi=30$,   where $\phi$ is the number of data points used in each fit of the linear regression for estimating the Hurst exponent.

\subsection{Measuring Changes in Stock Market Behavior}\label{subsec:Construct the Bull Index}

We propose a Bullish Index that measures   changes in stock market behavior. In line with Auer \citep{auer2016performance}, we start from a Bullish ratio ($B_t$) as the ratio between the number of  daily positive returns in a $H>0.5$ regime and the total number of trading days in a given month $t$:

\begin{equation}\label{B ratio}
B_{t} =   [\frac{\# \{ r(d)>0, H>0.5 ;\; d=1,\dotsc,M \} { \# \{r(d); \; d=1,\dotsc,M\} }]_t}, \;\; t=1,\dotsc,T
\end{equation} 

where $r(d)$ is the daily log return in month $t$; $M$ denotes the total number of trading days in a given month $t$, and  $T$  refers to the last observation   month.  In so doing,   calculating the ratio between the number of positive daily returns (for $H>0.5$, in this paper) to the total number of trading days in a given month, the Bullish ratio  becomes a monthly indicator. % point.

A Bullish Index ($BU_t$) is developed to measure the changes in the monthly Bullish ratio, as follows

\begin{equation}\label{Bull Index}
BU_{t}=log(\frac{B_{t}}{B_{t-1}}), \;\; t=2,\dotsc,T
\end{equation}

where $BU_{t}$ is not defined when $B_{t-1}=0$. (We replace $BU_{t}$ with zero in such a scenario for any $t=2,\dotsc,T$.)  A  positive persistent trend is captured if  $BU_t>0$. A   negative persistent trend  is given when $BU_t \leq 0$.

The merit of calculating the Bullish index instead of using the $B_t$ ratio is that the Bullish index can better capture changes in stock market behavior. For example, when $B_{t}=20$ increases to $B_{t+1}=21$,  the Bullish index would   be considered  as a "marked change of stock market behavior", while a high   $ B_t$ ratio ``simply''  implies a strong positive persistent-like behavior. Thus, the "peak" and "trough" of the Bullish index   help  one to identify the unstable periods (of interest) in the market.

\section{Data and Summary Statistics}\label{sec:datasum}
\subsection{Local Hurst Exponent and Bullish Index}\label{sec:Local Hurst Exponent and Bull Index}
In view of   constructing  the Bullish Index, the daily closing prices of the S\&P500 Index are obtained from the Institutional Broker's Estimate System (IBES).

\medskip
\begin{center}[Insert Table \ref{tab:summary stat table} about here]\end{center} 
\medskip

The sampling period runs from 1950:12:01 till 2019:12:31. Table \ref{tab:summary stat table} presents the mean, standard deviation (Std. Dev.), minimum (min.), maximum (max.), and the first-order autocorrelation coefficient of the daily log returns, the local Hurst exponent, the Bullish Index and the Bullish ratio. The local Hurst exponent ranges from 0.196 to 0.786, with a mean of 0.557, and a standard deviation of 0.094. The monthly Bullish Index ranges from -2.848 to 2.660, with mean and standard deviation equal to -0.427 and 0.479\%, respectively. 

\begin{figure}[th]
	\centering

	\setlength{\fboxrule}{0pt}
	\caption{Time-varying Local Hurst exponent of S\&P500 Index (1950-2019)
	}
	\caption*{Plot of the time series of local Hurst exponent of S\&P500 Index from 1950:12:01 to 2019:12:31. The gray areas denote NBER recessions.}
	{%        
		\fbox{\hspace*{-1.5cm} \includegraphics[scale=0.45]{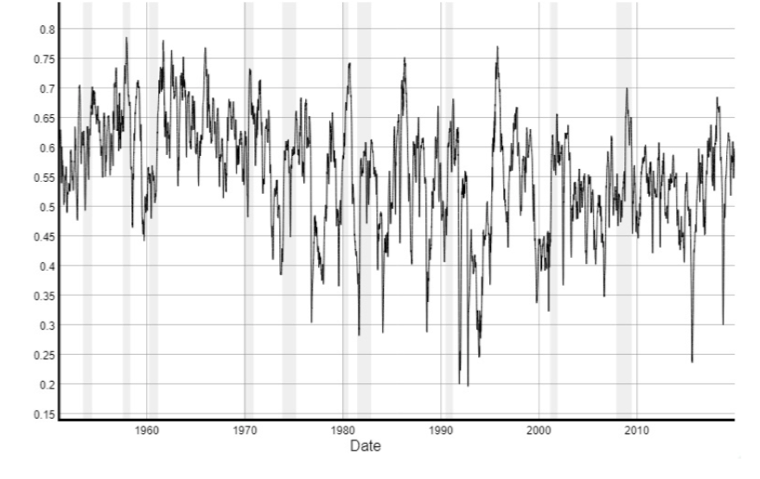}}}
	\label{tab:Local Hurst exponent}
\end{figure}

\medskip
\begin{center}[Insert Figure \ref{tab:Local Hurst exponent} about here]\end{center} 
\medskip

Figure \ref{tab:Local Hurst exponent} presents the plots of  the daily time-varying local Hurst exponent calculated with the  FDMAA  method from 1950:12:01 to 2019:12:31 for the S\&P500 Index. The grey areas denote the National Bureau of Economic Research (NBER) dated business-cycle recessions ($\href{https://www.nber.org/cycles.html}{https://www.nber.org/cycles.html}$). The decline in the local Hurst exponent implies that price movements become less predictable and indicate higher market efficiency \footnote{Alvarez-Ramirez et al. \citep{alvarez2008time} suggest that the end of the Bretton Woods system (BWS) in 1971-1972 had a significant impact on the functioning of the US stock markets. New agents were introduced into the stock market as the currencies were allowed to float freely, with some of them related to the ability to exploit differences and arbitrage in currencies exchange rates. In addition, Barunik and Kristoufek \citep{barunik2010hurst}  suggest that the introduction of algorithmic	trading, with exponentially rising trading volumes, is the main reason behind the value evolution of  the local Hurst exponent.}. 

\begin{figure}[th]
	\centering
	
	\setlength{\fboxrule}{0pt}
	    \caption{Time-varying Bullish Index (1950-2019)
	    }
    \caption*{Plot of the time series of Bullish Index from 1950:12:01 to 2019:12:31. The gray areas denote NBER recessions.}
    {%        
		\fbox{\hspace*{-1.5cm} \includegraphics[scale=0.45]{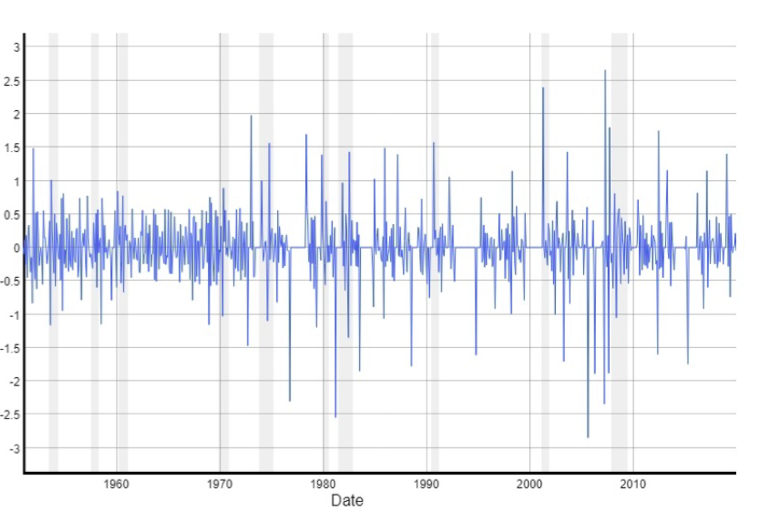}}}
	\label{tab:Time varying Bull Index}
\end{figure}

\medskip
\begin{center}[Insert Figure \ref{tab:Time varying Bull Index} about here]\end{center} 
\medskip

Figure \ref{tab:Time varying Bull Index} presents the plots of  the monthly time-varying Bullish Index. We define a peak (trough) as the top (bottom) 2.5\% quantile of the Bullish Index, with $BU_t>1$ ($BU_t<-1$) \footnote{We set the threshold at the 2.5\% level to ensure that the changes in Bullish ratio are substantial enough for consideration; we have performed four robustness checks to validate the empirical results; see Section \ref{subsec:Robustness check}.   }. The peak (trough) of the Bullish Index is observed when there is a massive increase (decrease) in the number of positive returns with $H>0.5$ in a trading month.
% \footnote{ If the number of positive returns with $H>0.5$ in a trading month is gradually increasing, the Bullish Index would have a small positive value.}. 
 Positive shocks are observed prior to or at the very beginning of recessions (1953, 1973, 1980, 1990, 2001 and 2007). The three largest values of the Bullish Index are recorded in 2007:04 (the Great Recession), 2001:04 (Dot-com bubble), and 1973:01 (the oil crisis), with $BU_t$=2.660, 2.398 and 1.979, respectively, which is associated with the most severe recessions in the history. The three smallest values of the Bullish Index are recorded in 2005:08, 1981:03 (the international debt crisis), and 2007:03 (the Great Recession), with $BU_t$=-2.848, -2.544 and -2.344, respectively, which are also related to recessions.

\subsection{Predictors}\label{subsec:Predictors}

Recall that in this paper, we focus on the impact of changes in stock market behavior on equity risk premium predictability regarding macroeconomic variables and technical indicators. The macroeconomic data, monthly S\&P500 Index and risk-free rate are obtained from the webpage $\href{http://www.hec.unil.ch/agoyal/}{http://www.hec.unil.ch/agoyal/}$. The (14) macroeconomic variables used in this study have been widely adopted in academic studies \vphantom{\citep{ren2019balanced,wang2019oil,zhang2019forecasting}}\citep{ren2019balanced}-\citep{zhang2019forecasting}, see details in Appendix \ref{sec:Description of the 14 macroeconomic variables} \footnote{More details on the possible and optimized choices of macroeconomic variables can be found in Welch and Goyal \citep{welch2008comprehensive}.}.

\begin{center}[Insert Table \ref{tab:summary stat table macro} about here]\end{center}

Table \ref{tab:summary stat table macro} reports the summary statistics of the log equity risk premium and of   14 macroeconomic variables from 1950:12:01 till 2019:12:31. \footnote{Note that the Mele (asymmetric)   measure is used to estimate the equity risk premium volatility (ERPV) \citep{mele2007asymmetric}.}

As in Neely et al. \citep{neely2014forecasting}, we construct the (14) technical indicators based on three popular trend-following strategies: moving average rule, momentum rule and on-balance volume, see details on notations in Appendix \ref{sec:Description of the technical indicator}.

\section{Empirical Results}\label{sec:Empirical Results}
\subsection{In-Sample Analysis}\label{subsec:insampleanalysis}
In this section, we analyze the in-sample predictability of equity risk premium through the   ordinary least square (OLS) model:

\begin{equation}\label{tab:regressionecon}
r^{(i)}_{t+1}=\alpha^{(i)}+\beta^{(i)} \, \psi^{(i)}_{t}+\varepsilon^{(i)}_{t+1} \;\; 
\end{equation}
 where  $r^{(i)}_{t+1}$\footnote{Recall that we denote $r(d)$ as the daily log returns and $r_{t}$ as the monthly equity risk premium.} is  the monthly equity risk premium, i.e., the return on the S\&P500 Index in excess of the risk-free rate from
period $t$ to $t+1$; this leads to 14 regression equations, for   each predictor $\psi^{(i)}$, since
 $i=1,...,14$;  $i=1$ refers to the dividend  to price ratio (D/P), $i=2$ refers to dividend yield (D/Y), etc., as listed  in the  Appendix \ref{Predictors} on  Macroeconomic Variables, 
%, while $R^{2}$ is the log equity premium start from 192701, which is a fixed column of data, so there is no $i$ for it, same applies for technical indicators.
% $\psi^{(i)}_{t}$ is the predictor at time $t$; 
 %$\psi^{(i)}_{t}$  takes the value of {\bf  [ I don't understand the graph; how many points are fitted by a straight line ?]} a macroeconomic variable, or a trading signal generated by a technical indicator; 
 $\alpha^{(i)}$ is a constant, $\beta^{(i)}$ is the slope coefficient, and $\varepsilon^{(i)}_{t+1}$ is a zero-mean disturbance term.  We test the null hypothesis of no predictability $H_{0}:\beta^{(i)}=0$ against $H_{1}: \beta^{(i)}>0$,
 % {\bf [ WHY IS   $\beta^{(i)}$ SURELY POSITIVE ? ] } {\color{black}  \bf A one-tailed test is appropriate if one only want to determine if there is a difference between groups in a specific direction, we are only interested in predictors with forecasting power. }
  using a heteroskedasticity-consistent
t-statistic for the sampling period [1950:12:01 - 2019:12:31]. 

To incorporate the information from multiple predictors, we follow the literature \citep{zhang2019forecasting,neely2014forecasting,sun2021can} and employ a principal component ($PC$) predictive regression model \footnote{Principal components parsimoniously incorporate information from a large number of potential predictors in a predictive regression. The first few principal components	identify the key comovements among the entire set of predictors, which filters out much of the noise in individual predictors, thereby guarding against in-sample overfitting. In this way, the overall performance of a set 	of predictors can be clearly observed whence compared to each other. } (PC-ECON, PC-TECH or PC-ALL model):

\begin{equation}\label{tab:PCA}
r^{(j)}_{t+1}=\alpha^{(j)}+\sum_{k=1}^{K}\gamma_{k} \, \hat{F}^{(j)}_{k,t}+\varepsilon^{(j)}_{t+1},\;\;  \textrm{for}\;\;  j= ECON, \;TECH, \;ALL.
\end{equation}
 For example,  $\hat{F}^{ECON}_{t}= (\hat{F}^{ECON}_{1,t},\dotsc,\hat{F}^{ECON}_{K,t})$ denotes the vector containing the first $K$ principal components estimated from the 14 \underline{macroeconomic variables}. Similarly,  $\hat{F}^{TECH}_{t}= (\hat{F}^{TECH}_{1,t},\dotsc,\hat{F}^{TECH}_{K,t})$ denotes the vector containing the first $K$ principal components estimated from all of the 14 \underline{technical indicators}; $\hat{F}^{ALL}_{t}= (\hat{F}^{ALL}_{1,t},\dotsc,\hat{F}^{ALL}_{K,t})$ denotes  the vector containing the first $K$ principal components estimated from all  (28) predictors taken together.
 
In practice, one can aggregate the 14 variables by considering more than one principal component; in such a case $K$ can be equal to 2, 3, 4,...,etc.  In the present study,  in line with the literature \citep{neely2014forecasting,sun2021can}, we need only to consider a maximum value  for $K$; in fact, $K $=3, here,  thereafter selecting the appropriate $K\le3$ such that the model has the highest regression coefficient $R^2$.

Finally, we estimate each model Eq.(\ref{tab:PCA}) via OLS, compute heteroskedasticity-consistent t-statistics, and base inferences on wild bootstrapped p-values. 
 Thereafter, the model with the highest adjusted $R^2$ is selected for discussion.

Table \ref{tab:In-sampleResults} reports the estimation results for in-sample analysis. Campbell and Thompson \citep{campbell2008predicting} suggest that a 0.5\%  monthly $R^{2}$   can represent an economically significant degree of predictability. As shown in the second column of Panel A, 6 of the 14 macroeconomic variables exhibit a quite  significant predictability with $R^{2}$ ranging from 0.36\% till 0.91\% for the entire sampling period, i.e.,  from 1951:01 till  2019:12;  the average $R^{2}$ of the 14 macroeconomic variables is 0.31\%.  In comparison, 12 of the 14 technical indicators exhibit a significant predictability with $R^{2}$ ranging  from 0.24\% to 0.93\%, as shown in the eleventh column of Panel A; the average $R^{2}$ for these 14 technical indicators regressions $\simeq$ 0.49\%.

To assess the relative strength of an equity risk premium predictability within the listed National Bureau of Economic Research (NBER)  business-cycle expansions or recessions, we follow Neely et al. \citep{neely2014forecasting} to compute the $R^{2}$ statistics for expansions or recessions. Neely et al. \citep{neely2014forecasting} suggest  to "compute the following intuitive versions of the conventional $R^2$ statistic":

\begin{equation}\label{tab:rinsample}
R_{\zeta}^{2}=1- \frac{\sum_{t=n}^{T} I_{t}^{\zeta}(r_{t}-\widehat{r}_{t})^{2}}{\sum_{t=n}^{T}I_{t}^{\zeta}(r_{t}-\bar{r}_{t})^{2}}, \; \textrm{for} \; \zeta= exp, \; rec,
\end{equation}

where $I_{t}^{exp}$  is a dummy variable that takes a value of 1 when month $t$ is $during$ an expansion  (bull market)  and 0 otherwise, while $I_{t}^{rec}$ takes a value of 1 when month $t$ is in a recession and 0  otherwise. The fitted value based on predictive regression Eq.(\ref{tab:regressionecon}) or Eq.(\ref{tab:PCA}) is denoted as $\widehat{r}_{t}$; $\bar{r}_{t}$ is the full-sample mean of equity risk premium; $T$ is the number of observations of the full-sampling period.
Notice that like for the whole sample $R^{2}$, the $R_{exp}^{2}$ and $R_{rec}^{2}$ statistics can be negative. 

As shown in the fourth and fifth columns of Table \ref{tab:In-sampleResults}, the equity risk premium predictability is substantially stronger for 10 of the 14 macroeconomic variables in recessions than in expansions. Similarly, all individual technical indicators also show stronger equity risk premium predictability in recessions than in expansions periods. 
 
 %Note that it has the same logic as expansion periods and recessions periods.
We are interested in whether equity risk premium predictability based on predictive regression Eq.(\ref{tab:regressionecon}) or Eq.(\ref{tab:PCA}) is influenced by the peak $(BU^{+})$ or trough $(BU^{-})$ of the Bullish Index. As in Eq.(\ref{tab:rinsample}), we compute the $R^{2}$ statistics for the peak or trough dummy by "replacing" $\zeta$ by $\eta$ as follows: for  $\eta= BU^{+}, \; stable^{+},  BU^{-}, \; stable^{-}$, we define $I_{t}^{BU^{+}}$ ($I_{t}^{BU^{-}}$) as the peak (trough) dummy by considering $BU^{+}_{t-3 \rightarrow t+3}$ ($BU^{-}_{t-3 \rightarrow t+3}$) which takes a value of 1 from three months before up to three months subsequently after the Bullish Index is greater (smaller) than its top (bottom) 2.5\% quantile level, $BU>1$ ($BU<-1$), take zero otherwise as the stable dummy, $I_{t}^{stable^{+}}$ ($I_{t}^{stable^{-}}$). We interpret the peak (trough) periods as the months when $I_{t}^{BU^{+}}=1$ ($I_{t}^{BU^{-}}=1$); $stable^{+}$ ($stable^{-}$) periods as the months when $I_{t}^{stable^{+}}=1$ ($I_{t}^{stable^{-}}=1$). \footnote{Note that the peak and trough periods can overlap when peak and trough signals are occurring closely. For example, $BU>1$ ($BU<-1$) is observed on 2007:04 (2007:03), the corresponding peak (trough) periods are 2007:01 to 2007:07 (2006:12 to 2007:06). In this way, we can observe the impact of positive shocks or negative shocks on equity risk premium predictability independently. }

\medskip
\begin{center}[Insert Table \ref{tab:In-sampleResults} about here]\end{center} 
\medskip

As shown in the sixth and seventh columns of Table \ref{tab:In-sampleResults} in Panel A, the equity risk premium predictability is stronger for 8 of the 14 macroeconomic variables in Bullish Index peak periods than in $stable^{+}$ periods, which includes: D/E, ERPV, B/M, NEER, LTR, TMS, DYS, and INFL. 

The fifteenth and sixteenth columns of Table \ref{tab:In-sampleResults} in Panel A show that 12 of the 14 technical indicators have higher equity risk premium predictability in Bullish Index peak periods than in $stable^{+}$ periods, which includes: MA(1,12), MA(2,9), MA(2,12), MA(3,9), MA(3,12), MOM(9), MOM(12), VOL(1,9), VOL(1,12), VOL(2,9), and VOL(3,9). 

In the last two columns of Table \ref{tab:In-sampleResults} in Panel A, the results show that all individual technical indicators have higher equity risk premium predictability in trough periods than in $stable^{-}$ periods:    1.18, 2.10, 1.57, 2.71, etc. are. all greater than 0.22, 0.31, 0.30, 0.51, etc., respectively. There is no such evidence found for macroeconomic variables.

In Panel B of Table \ref{tab:In-sampleResults}, the PC-ECON model   $BU^{+}$ shows substantially stronger equity risk premium predictability in peak periods than in $stable^{+}$ periods,  with a $R^{2}=$  1.35\% and 0.49\%, respectively \footnote{The PC-ECON, PC-TECH and PC-ALL models have a $K = 1$, as chosen from the adjusted $R^2$. }. In contrast, the PC-TECH model shows stronger equity risk premium predictability from the  Bullish Index trough periods ($BU^{-}$)  than in $stable^{-}$ periods with a $R^{2}$ of 1.82\% and 0.23\%, respectively. 

In Panel C of Table \ref{tab:In-sampleResults}, the PC-ALL model shows strong equity risk premium predictability in peak periods ($R^{2}=2.72\%$), trough periods ($R^{2}=3.13\%$), and recessions ($R^{2}=5.04\%$). 

Thus,   these empirical results imply that the peak and trough of the Bullish Index contain valuable information on equity risk premium predictability.% In particular, peak periods tend to be associated with stronger equity risk premium predictability for forecasts based on macroeconomic variables.%  {\bf [ that is far from obvious! ]} {\color{black}  \bf 9.64\% for PC model is written out now.} 

\subsection{Out-of-sample Analysis}\label{subsec:outofsampleanalysis}
In this section, we investigate the impact of changes in stock market behavior on equity risk premium predictability subsequently after the peaks and troughs of the Bullish Index. To avoid look-ahead bias, we start with an initial estimation window from 1950:12:01 till 1965:12:31 to estimate predictive regression Eq.(\ref{tab:oosregression}) to evaluate the first out-of-sample forecast of equity risk premium in January 1966. 

We recursively estimate regression Eq.(\ref{tab:oosregression}) to evaluate the one-step-ahead forecast of equity risk premium and repeat these steps until the end of the sampling period; similarly to the above,

\begin{equation}\label{tab:oosregression}
\hat{r}^{(i)}_{t+1}=\hat{\alpha}^{(i)}_{t}+\hat{\beta}^{(i)}_{t} \, \psi^{(i)}_{t},
\end{equation}

where $\hat{r}^{(i)}_{t+1}$ is the one-step-ahead forecast of equity risk premium; $ \psi^{(i)}_{t}$ is the level of a macroeconomic variable, or the trading signal generated by a technical indicator. For each time step $t$,  $\hat{\alpha}_{t}$ and $\hat{\beta}_{t}$ are obtained by regressing the realized equity risk premium series $\{r^{(i)}_{s} \}_{s=2}^{t}$ on a constant $\hat{\alpha}^{(i)}_{t}$  and on the lagged predictors $\{\psi^{(i)}_{s} \}_{s=1}^{t-1}$ using the data available up to month $t$. We also estimate the out-of-sample forecasts based on principal component predictive regression,  
\begin{equation}\label{tab:oosregressionpca}
\hat{r}^{(j)}_{t+1}=\hat{\alpha}^{(j)}_{t}+\sum_{k=1}^{K}\hat{\gamma}_{t,k} \, \hat{F}_{1:t,k,t}^{(j)}, \;\; \textrm{for} \; j=ECON, \; TECH, \;\textrm{or} \; ALL;
\end{equation} 

where $\hat{F}_{1:t,k,t}^{(j)}$ is the $k$-th principal component  
 extracted from the 14 macroeconomic variables $(j=ECON)$, 14 technical indicators $(j=TECH)$, or all of the 14 macroeconomic variables and 14 technical indicators taken together $(j=ALL)$ estimated using the data
through $t$. As in section \ref{subsec:insampleanalysis}, we run model Eq.(\ref{tab:oosregressionpca}) using different value of $K$ with $ K \leq 3$. Next,  the model with the highest adjusted $R^2$ is selected; 
$\hat{\alpha}_{t}$ and $\hat{\gamma}_{t,k}$ $(k = 1,\dotsc,K)$ are the OLS estimates from recursively regressing the realized equity risk premium series $\{r_{s} \}_{s=2}^{t}$ on a constant $\hat{\alpha}^{(j)}_{t}$ and the lagged principal component(s) $\{\hat{F}_{1:t,k,t}^{(j)} \}_{s=1}^{t-1}$ $(k = 1,\dotsc,K)$.

Welch and Goyal \citep{welch2008comprehensive} show that predictive regression based on individual macroeconomic variables typically underperforms historical average (HA) returns forecast in out-of-sample analysis. In the present study, 
by performing similar tests, with our concerns on peak and trough, we can see whether macroeconomic variables can outperform the historical benchmark in the peak periods; in other words, 
 we can see how bad   the macroeconomic variables are as compared to a  HA and  how useful  the macroeconomic variables can be   compared to HA in peak periods.
 
  The historical average (HA) forecast is given by

\begin{equation}\label{tab:HA}
\widehat{r}_{t+1}^{HA}=\frac{1}{t}\sum_{\tau=1}^{t} r_{\tau}.
\end{equation}

where $\widehat{r}_{t+1}^{HA}$ is the monthly equity risk premium forecast by the historical average equity risk premium. We employ the Campbell and Thompson \citep{campbell2008predicting} $R_{OS}^{2}$ to evaluate the out-of-sample performance of predictive regression Eq.(\ref{tab:oosregression}) and Eq.(\ref{tab:oosregressionpca}). Their $R_{OS}^{2}$ statistic is defined as:

\begin{equation}\label{tab:roos}
R_{OS,\zeta}^{2}=1-\frac{\sum_{t=n}^{T-1}I_{t}^{\zeta}(r_{t+1}-\widehat{r}_{t+1})^{2}}{\sum_{t=n}^{T-1}I_{t}^{\zeta}(r_{t+1}-\widehat{r}^{HA}_{t+1})^{2}}, \; \textrm{for} \; \zeta= entire, \; exp, \; rec.
\end{equation}

\noindent where $r_{t+1}$ is the actual monthly equity risk premium at time $t+1$, $\widehat{r}_{t+1}$ is the forecasted equity risk premium from predictive regression Eq.(\ref{tab:oosregression}) or Eq.(\ref{tab:oosregressionpca}), and $\widehat{r}^{HA}_{t+1}$ is the historical average benchmark. We denote $I_{t}^{entire}$ as the entire sampling periods; $I_{t}^{exp}$ ($I_{t}^{rec}$) is a dummy variable that takes a value of 1 when month $t$ is an expansion (recession) and zero otherwise. The $R_{OS}^{2}$ statistics lie in the range $(-\infty,1\rbrack$. The predictive regression forecast $\widehat{r}_{t+1}$ has a lower mean squared forecasting error (MSFE) than the historical average $\widehat{r}^{HA}_{t+1}$ when $R_{OS}^{2}>0$. We employ the Clark and West \citep{clark2007approximately}  MSFE-adjusted statistics  to test the hypothesis $H_{0}: R_{OS}^{2}\leq0$ against $H_{1}: R_{OS}^{2}>0$.

\medskip
\begin{center}[Insert Table \ref{tab:oosResults} about here]\end{center} 
\medskip

Panel A of Table \ref{tab:oosResults} reports the out-of-sample forecasting results based on either  individual macroeconomic variables or technical indicators from 1966:01:03 to 2019:12:31. For the entire sampling period, among the 14 macroeconomic economic variables, only LTR has outperformed the historical average benchmark with a positive $R_{OS}^{2}= 0.49\%$, see  the third column of Panel A. In contrast, 11 of the 14 technical indicators have a positive $R_{OS}^{2}$ range from 0.09\% to 0.67\% : see the twelfth column of Panel A. 

From the recession/expansion point of view, 8 macroeconomic variables exhibit stronger equity risk premium predictability in recessions than in expansions as shown in the fourth and fifth columns of Panel A. In the thirteenth and fourteenth columns of Panel A,  all the individual technical indicators indicate a higher equity risk premium predictability in recessions than in expansions. 

As in Section \ref{subsec:insampleanalysis}, by replacing $\zeta$ as $\eta$ in Eq.(\ref{tab:roos}), for  $\eta= BU^{+}, \; stable^{+},  BU^{-}, \; stable^{-}$, we compute the $R_{OS}^{2}$ statistics for the peak (trough) dummy, $I_{t}^{BU^{+}}$ ($I_{t}^{BU^{-}}$) by considering $BU^{+}_{t+1 \rightarrow t+3}$ ($BU^{-}_{t+1 \rightarrow t+3}$) which takes a value of 1 for the three months subsequently after the Bullish Index is greater (smaller) than its top (bottom) 2.5\% quantile level, $BU>1$ ($BU<-1$) at time $t$, and take zero otherwise as the stable dummy, $I_{t}^{stable^{+}}$ ($I_{t}^{stable^{-}}$). We interpret peak periods ($BU^{+}$) and trough periods ($BU^{-}$) as the months when $I_{t}^{BU^{+}}=1$ and $I_{t}^{BU^{-}}=1$, respectively.

In the sixth and seventh columns of Panel A in Table \ref{tab:oosResults}, 9 macroeconomic variables exhibit stronger predictive power in peak periods than in $stable^{+}$ periods. While only three technical indicators: MA(1,9), MA(1,12), and MA(2,9), have a $R_{OS}^{2}$ statistics with 1.44\%, 0.92\%, and 0.81\%, respectively, higher than $stable^{+}$ periods as shown in the fifteenth and sixteenth columns in Panel A. 

In contrast, all individual technical indicators have a $R_{OS}^{2}$ statistics range from 0.37\% to 2.31\% in trough periods, which is higher than $stable^{-}$ periods as shown in the last two columns of Panel A. Whereas only six macroeconomic variables have a $R_{OS}^{2}$ statistics range from 0.93\% to 4.50\% in trough periods, higher than the $stable^{-}$ periods as shown in the eighth and ninth columns in Panel A.

Panel B of Table \ref{tab:oosResults} reports out-of-sample results of the PC-ECON and PC-TECH forecasts.
The results show that the PC-TECH forecast outperforms both PC-ECON forecast and HA forecast with a $R_{OS}^{2}= 0.44\%$ for the entire sampling period. The out-of-sample predictive ability of both PC-ECON forecast and PC-TECH forecast are stronger in recessions than in expansions. 

More importantly, the results show a substantial $R_{OS}^{2}= 9.64\%$ for the PC-ECON model in the peak periods, and a $R_{OS}^{2}= 2.10\%$ for PC-TECH model in the trough periods, which is considerably higher than the corresponding stable periods. Moreover, the $R_{OS}^{2}$’s are statistically significant at the 1\% level, according to the MSFE-adjusted statistics \footnote{The MSFE-adjusted statistics are not reported in the table for subperiods to avoid an overfull table.}, suggesting
that the MSFE of out-of-sample forecasts generated by PC-ECON (PC-TECH) model in the peak (trough) periods is significantly lower than that of the historical average.

Panel C of Table \ref{tab:oosResults} reports out-of-sample results for the PC-ALL forecast. The results suggest strong equity risk premium predictability in recessions ($R_{OS}^{2}= 11.26\%$), Bullish Index peak periods ($R_{OS}^{2}= 11.89\%$) and trough periods ($R_{OS}^{2}= 5.42\%$), respectively; note that the $R_{OS}^{2}$’s are statistically significant at the 1\% level.

Thus,  in summary, we find a strong equity risk premium predictability with macroeconomic variables, especially LTR, TMS, and D/P for up to three months after a positive shock in the market. Moreover,  technical indicators exhibit strong equity risk premium predictability after a negative shock. 

\subsection{Asset Allocation Analysis}\label{subsec:assetallocation}
We further evaluate the potential economic value from an asset allocation perspective. Following the literature \citep{campbell2008predicting,neely2014forecasting,ferreira2011forecasting}, we consider a risk-averse investor with a relative risk-aversion coefficient of five who optimally allocates across equities and risk-free bills using either the historical average forecast Eq.(\ref{tab:HA}) or the predictive regressions Eq.(\ref{tab:oosregression}, \ref{tab:oosregressionpca}). At the end of month $t$, the investor rebalances the portfolio and allocates the following weights of equities in the portfolio:

\begin{equation}\label{tab:markowitz}
w_{t}=\frac{1}{\kappa} \frac{\widehat{r}_{t+1}}{\widehat{\sigma}^{2}_{t+1}}
\end{equation}

\noindent where $\kappa$ is the risk-aversion coefficient; $\widehat{r}_{t+1}$ is the one-step-ahead forecast of equity risk premium using either the historical average forecast Eq.(\ref{tab:HA}) or the predictive regressions Eq.(\ref{tab:oosregression}, \ref{tab:oosregressionpca}), $\widehat{\sigma}^{2}_{t+1}$ is the forecast of corresponding variance \footnote{As in Campbell and Thompson \citep{campbell2008predicting}, we assume that the investor uses a five-year moving window of past monthly returns to estimate $\widehat{\sigma}^{2}_{t+1}$.}. The risk-averse investor allocates $1-w_{t}$ of the portfolio in risk-free bills, and the realized portfolio return at the end of each period is calculated as:

\begin{equation}\label{tab:markowitzport}
{r}_{t+1}^{p}= w_{t}{r}_{t+1}+(1-w_{t}){r}_{t+1}^{f}
\end{equation}

\noindent where ${r}_{t+1}^{f}$ is the risk-free return. We impose some restriction on $w_{t}$ in order to exclude short sales and to have at most 50\% leverage; $w_{t}$ values lie in the range $[0 \; ; \; 1.5]$.  We calculate the "certainty equivalent return" (CER) of the portfolios as follows:

\begin{equation}\label{tab:cer}
{CER}_{p}= \bar{r}_{p}-\frac{\kappa}{2}{\sigma}_{p}^{2}
\end{equation}

\noindent where $\bar{r}_{p}$ and ${\sigma}_{p}^{2}$ are the mean and variance, respectively; ${CER}_{p}$ is the $CER$ of the portfolios with weights, $w_{t}$ in equities and $1-w_{t}$ in risk-free bills. The $CER$ can be regarded as the portfolio management fee that the investor would be willing to pay to exploit the information in each predictive model. 

Moreover consider the certainty equivalent return gain ($CER_{g}$) as the difference between the $CER$ for the investor who uses predictive regression forecast of equity risk premium based on Eq.(\ref{tab:oosregression}) or Eq.(\ref{tab:oosregressionpca}) and the $CER$ for an investor who uses the historical average forecast given by Eq.(\ref{tab:HA}). We multiply this difference by 12 such that the $CER_{g}$ can be regarded as the annual portfolio management fee that an investor would be willing to pay to access the predictive model.
%\hspace*{-1.5cm}\resizebox{26cm}
% \usepackage{multirow}
% \usepackage{graphicx}
% \usepackage{lscape}

%\hspace*{-1.5cm}\resizebox{26cm}
% \usepackage{multirow}
% \usepackage{graphicx}
% \usepackage{lscape}
% \usepackage{multirow}
% \usepackage{graphicx}
% \usepackage{lscape}

% \usepackage{graphicx}
% \usepackage{multirow}
% \usepackage{graphicx}
% \usepackage{lscape}
% \usepackage{multirow}  \hspace*{-1.5cm}\resizebox{26cm}
% \usepackage{graphicx}
% \usepackage{lscape}
% \usepackage{multirow}
% \usepackage{graphicx}
% \usepackage{lscape}
\medskip
\begin{center}[Insert Table \ref{tab:assetallocation} about here]\end{center} 
\medskip

% \usepackage{multirow}
% \usepackage{multirow}
% \usepackage{graphicx}
% \usepackage{multirow}
% \usepackage{graphicx}
% \usepackage{multirow}
% \usepackage{graphicx}
% \usepackage{multirow}
% \usepackage{graphicx}
% \usepackage{multirow}
% \usepackage{graphicx}
% \usepackage{multirow}
% \usepackage{graphicx}
% \usepackage{multirow}
% \usepackage{graphicx}
% \usepackage{multirow}
% \usepackage{graphicx}
% \usepackage{multirow}
% \usepackage{graphicx}
% \usepackage{multirow}
% \usepackage{graphicx}

For the entire sampling period as shown in the second column of Panel A in Table \ref{tab:assetallocation}, eight macroeconomic variables have a positive $CER_{g}$ which ranges from 0.18\% to 1.82\%. In the tenth column of Panel A, all of the technical indicators have a positive $CER_{g}$ ranging from 0.39\% to 2.77\%. The $CER_{g}$ is substantially larger in recessions than expansions for the majority of macroeconomic variables and for all   technical indicators. The $CER_{g}$ for 9 technical indicators outperforms macroeconomic variables.

We compute the $CER_{g}$ for the Bullish Index peak (trough) dummy, $I_{t}^{BU^{+}}$ ($I_{t}^{BU^{-}}$) and the stable dummy, $I_{t}^{stable^{+}}$ ($I_{t}^{stable^{-}}$). Consistent with the  statistical values reported in Table \ref{tab:oosResults}, the $CER_{g}$ is substantially larger for 9 of the 14 macroeconomic variables in peak periods than in the $stable^{+}$ periods, as shown in the fifth and sixth columns of Panel A in Table \ref{tab:assetallocation}. There is no significant impact on the $CER_{g}$ for technical indicators in the peak periods.

Almost all of the technical indicators have a larger $CER_{g}$ in trough periods than in the $stable^{-}$ periods as shown in the last two columns of Panel A in Table \ref{tab:assetallocation}. No similar evidence is found for macroeconomic variables in trough periods.

Panel B and C of Table \ref{tab:assetallocation} report the portfolio performance measures for the PC-ECON, PC-TECH and PC-ALL models. The results show that both PC-ECON and PC-TECH models have a substantially larger $CER_{g}$ in recessions than expansions. More importantly, the result in Panel B shows a substantial $CER_{g}$ of 7.21\% (6.30\%) for the PC-ECON (PC-TECH) model in peak (trough) periods which is considerably larger than the $stable^{+}$ ($stable^{-}$) periods of 1.03\% (1.47\%) in the fifth and sixth (last two) columns.

The PC-ALL model shows a substantially larger $CER_{g}$ in both peak and trough periods than the $stable^{\pm}$ periods in the sixth to ninth columns of Panel C in Table \ref{tab:assetallocation}. As a robustness check, we further conduct the asset allocation analysis with a risk-aversion coefficient of 1 or 3. Consistent with the findings in Table \ref{tab:assetallocation}, for a risk-aversion coefficient equal to 1 or 3, the PC-ECON model has a $CER_{g}$ of 2.74\% and 7.74\% in peak periods which is substantially higher than the $stable^{+}$ periods of -0.52\% and 0.79\%, respectively. Similarly, for risk-aversion coefficient equal to 1 or 3, the PC-TECH model has a $CER_{g}$ of 4.62\% and 7.31\% in trough periods which is substantially higher than the $stable^{-}$ periods of 0.11\% and 1.86\%, respectively. \footnote{Following Neely et al. \citep{neely2014forecasting}, Ferreira and Santa-Clara \citep{ferreira2011forecasting},  and many others, we have checked the robustness of the asset allocation performance analysis by imposing a proportional transaction cost of 50 basis points per transaction. For a risk-aversion coefficient equal to 5, the results show a substantial $CER_{g}$ of 6.97\% (5.55\%) for the PC-ECON (PC-TECH) model in peak (trough) periods which remains larger than the $stable^{+}$ ($stable^{-}$) periods of 0.15\% (1.07\%). The results remain robust even with such transactions costs. We have also obtained quantitatively similar results by considering a risk-averse investor with a relative risk-aversion coefficient equal to 1 or 3.} 

In summary, the results are robust across different levels of risk-aversion ($\kappa=1, 3$ or $5$), moreover even, taking   transaction costs into consideration.

\medskip
\begin{center}[Insert Table \ref{tab:holding periods} about here]\end{center} 
\medskip

The peak and trough periods in Table \ref{tab:assetallocation} focus on a three months investment horizon. We further investigate the impact of such shocks for different holding periods (investment horizon). Table \ref{tab:holding periods} reports the annualized certainty equivalent return gain (in percent) for different holding periods ($G_{t}$) subsequently after the Bullish Index reaches its peak (trough), $BU>1$ ($BU<-1$), with the consideration of proportional transactions cost of 50 basis points per transaction.

As shown in the third (eleventh) column of Table \ref{tab:holding periods}, the $CER_{g}$ remains at a high level of 15.56\% (9.85\%) from $t+4$ to $t+6$ subsequently after a positive shock (negative shock) for the PC-ECON (PC-TECH) model. The high equity risk premium predictability of the PC-ECON (PC-TECH) model is almost vanished from $t+7$ to $t+9$ ($t+10$ to $t+12$) after a positive shock (negative shock). In the seventh and last columns, the PC-ALL model have a $CER_{g}$ of 16.43\% and 16.85\% from $t+4$ to $t+6$ for the peak and trough periods, respectively.

\begin{figure}[th]
	\centering
	\setlength{\fboxrule}{0pt}
	\caption{Plot of certainty equivalent return gain ($CER_{g}$) for holding periods from one month to twelve months following a Bullish Index peak or trough.}
			\includegraphics[width=1\linewidth]{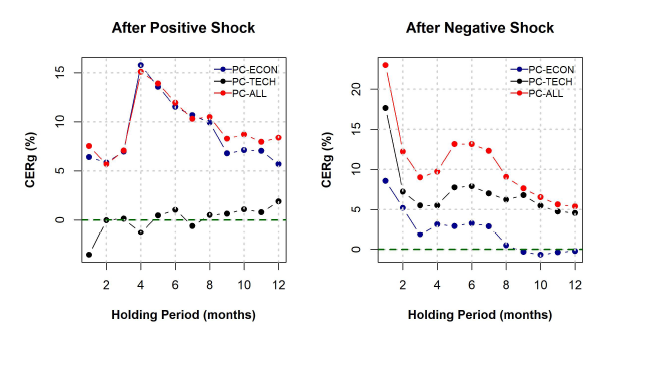}
	\label{fig:plotCER}
\end{figure}

\medskip
\begin{center}[Insert Figure \ref{fig:plotCER} about here]\end{center} 
\medskip

Figure \ref{fig:plotCER} presents the plots of the net-of-transactions-costs $CER_{g}$ for holding periods from one month to twelve months following Bullish Index peak or trough. After positive shocks, the $CER_{g}$ of the PC-ECON model reaches a maximum of 15.79\% for a four months holding periods and gradually decrease afterwards. The $CER_{g}$ of the PC-TECH model is close to zero and increases to 1.89\% for a twelve months holding periods.

Both PC-ECON and PC-TECH models have a substantial $CER_{g}$ one month subsequently after the negative shocks with 8.58\% and 17.68\%, respectively, and deteriorate rapidly but still being maintained at a high level. 

Thus,    the results imply that the PC-ECON (PC-TECH) model manifest stronger equity risk premium predictability than the historical average forecast for up to six (nine) months subsequently after the positive (negative) shocks. As shown in Figure \ref{fig:plotCER}, investors can exploit this information to yield an annualized certainty equivalent return gain of 11.96\% (13.16\%) in peak (trough) periods based on the PC-ALL model within six months following the positive (negative) shocks, even when a proportional transactions cost of 50 basis points per transaction is imposed.

\subsection{Robustness check}\label{subsec:Robustness check}
\subsubsection{Excluding recession dummy}\label{subsubsec:Excluding recession dummy}
The stronger equity risk premium predictability following a positive shock or a negative shock may simply due to its ability in detecting the recessions in advance as shown in Figure \ref{tab:Time varying Bull Index}.

% \usepackage{graphicx}
% \usepackage{multirow}
% \usepackage{graphicx}
% \usepackage{multirow}
% \usepackage{graphicx}
\medskip
\begin{center}[Insert Table \ref{tab:Robustness exclude recession} about here]\end{center} 
\medskip

For robustness check, we repeat the asset allocation analysis as in Table \ref{tab:assetallocation} by removing the recession dummy from the peak periods ($I_{t}^{BU^{+}}$) and trough periods ($I_{t}^{BU^{-}}$) dummy. The net-of-transactions-costs annualized certainty equivalent return gain ($CER_{g}$) in percent is reported Table \ref{tab:Robustness exclude recession}.

Panel B and C report the portfolio performance measures for PC-ECON, PC-TECH and PC-ALL models. After excluding recessions, the PC-ECON (PC-TECH)
model has a certainty equivalent return gain of 6.44\% (2.50\%) in peak (trough) periods, and 0.39\% (1.47\%) in the $stable^{+}$ ($stable^{-}$) periods. The results show stronger equity risk premium predictability following positive (negative) shocks with PC-ECON (PC-TECH) model remains robust even when recessions are excluded.

\subsubsection{Subsample period analysis}\label{subsubsec:Subperiod analysis}
For subsample period analysis,  we split the entire sampling period into two parts and repeat the asset allocation analysis as in Table \ref{tab:assetallocation} with a proportional transactions cost of 50 basis points per transaction. The first subsampling period spans from 1966:01:03 to 1993:12:31. The result shows that the PC-ECON (PC-TECH)
model has a certainty equivalent return gain of 10.67\% (2.38\%) in peak (trough) periods and 5.13\% (0.28\%) in the $stable^{+}$ ($stable^{-}$) periods. The second subsampling period spans from 1994:01:03 to 2019:12:31. The result shows that the PC-ECON (PC-TECH)
model has a certainty equivalent return gain of 2.98\% (8.96\%) in peak (trough) periods and -5.02\% (1.93\%) in the $stable^{+}$ ($stable^{-}$) periods. The $R_{OS}^{2}$’s of PC-ECON (PC-TECH) model are statistically significant at the 1\% level in peak (trough) periods in both subsamples, according to the MSFE-adjusted statistics.

Thus,   the findings are consistent for the entire sampling period: from the results in Table \ref{tab:assetallocation}, it is seen that the PC-ECON (PC-TECH) model has a high level of certainty equivalent return gain in peak (trough) periods across both of the subsample periods even when a proportional transaction cost of 50 basis points per transaction is imposed \footnote{The table is not reported here for saving space.}.

\subsubsection{Excluding outliers}\label{subsubsec:Excluding outlier}
The above analysis focuses on the extreme values of the Bullish Index which involves the analysis of possible outliers. In this section, we repeat the asset allocation analysis as in Table \ref{tab:assetallocation} by removing the top and bottom 5\% of the Bullish Index, and consider the top 5\% (bottom 5\%) of the trimmed Bullish Index as the new peak (trough) dummy. The results remain robust where the PC-ECON (PC-TECH)
model has a net-of-transactions-costs certainty equivalent return gain of 5.96\% (3.74\%) in peak (trough) periods, which remains at a higher level than for  the $stable^{+}$ ($stable^{-}$) periods when we obtain 0.11\% (1.21\%). The $R_{OS}^{2}$ of PC-ECON (PC-TECH) model are statistically significant at the 1\% level in the alternative peak (trough) periods, according to the MSFE-adjusted statistics. The findings are consistent with the results in Table \ref{tab:assetallocation}: the PC-ECON (PC-TECH) model exhibits strong equity risk premium predictability in the alternative peak (trough) periods.

\subsubsection{Without the consideration of Bullish ratio}\label{subsubsec:Without consideration of B ratio}
Moreover, Eom et al. \citep{eom2008hurst} show that a stock market Index with a high Hurst exponent tends to have a high hit rate, the "predictability",  with the nearest-neighbor method. Within our framework, it is of interest to calculate the monthly local Hurst exponent, i.e.,   the mean of the daily local Hurst exponent, and compute the MSFE for the HA forecast, the PC-ECON,  and PC-TECH forecasts for the periods when $H>0.5$ and $H \leq 0.5$. 
We find that the historical average forecast, PC-ECON and PC-TECH forecasts have a lower MSFE in the months when $H>0.5$ than $H \leq 0.5$; this is consistent with Eom et al. \citep{eom2008hurst}. 

We are also interested in whether the PC-ECON or PC-TECH forecast has a stronger equity risk premium predictability in the months when $H>0.5$ in outperforming the historical average benchmark than when $H \leq 0.5$, without the consideration of the Bullish ratio. 
The results show that the PC-ECON model has a $R_{OS}^{2}$ statistics of -3.27\% during the months when $H>0.5$ and 3.01\% when $H \leq 0.5$. Similarly, the PC-TECH model has a $R_{OS}^{2}$ statistics of 0.36\% in the months when $H>0.5$ and 0.59\% when $H \leq 0.5$. \footnote{The full results for individual macroeconomic variables and technical indicators are not reported here for saving space, we find no evidence of stronger equity risk premium predictability based on individual predictors for $H>0.5$ or $H>0.6$ in outperforming the historical average forecast.} The results suggest that forecasts based on the PC-ECON or PC-TECH model have no stronger equity risk premium predictability for $H>0.5$ than when $H \leq 0.5$ in outperforming the historical average benchmark. We obtain quantitatively similar results when using a more restrictive approach, like for $H>0.6$. This further supports the use of the Bullish ratio as a valuable complement to the local Hurst exponent.

\section{Conclusion}\label{sec:conclusion}
This paper reports some investigation indicating that a substantial change in stock market behavior has a statistically and economically significant impact on equity risk premium predictability through forecasts based on macroeconomic variables or technical indicators. 

We base our analysis on the construction of a so called Bullish Index, emanating from the Bullish ratio motivated by Auer \citep{auer2016performance}. Its local Hurst exponent  is studied classically. We are aware that not all variants are considered here. Nevertheless, even though the data is limited to  the US market, precisely to the S\&P500,  the various behaviors, recessions and expansions regimes, peaks and troughs, so examined, plus the various tests suggest that the conclusion can be qualitatively and even quantitatively similar for other matured markets. 

We make two major contributions to the existing literature. First, we indicate that the findings suggest that positive shocks (‘‘peaks’’) of the Bullish Index are  closely related to a strong equity risk premium predictability, when forecasts are based on macroeconomic variables,  outperforming historical average forecast, for up to six months following a positive shock. In contrast, a negative shock (a ‘‘trough’’) is associated with a strong and significant equity risk premium predictability   outperforming the historical average forecast,  for up to nine months following a negative shock, when the forecast is based on technical indicators. Notice that these empirical results suggest an evident violation of the EMH weak-form \citep{jovanovic2018comparison}. 

Therefore, from an asset allocation point of view, we propose that investors can benefit from  forecasts based on both macroeconomic variables and technical indicators (the here above called PC-ALL model).  For the examined data, the method yields an annualized certainty equivalent return gain of 11.96\% in peak periods and 13.16\% in trough periods for up to six months following the shocks (positive or negative shocks), even when a proportional transaction cost of 50 basis points per transaction is imposed. 
  
 We are aware that our conclusions might not be universal, but could be challenged on a numerical point of view, in particular if the method is applied to   financial indices other than the S\&P500, on other time intervals, and on other financial indicators. Moreover, one can study cases in which $H<0.5$, and/or one counts daily negative returns, and various combinations of such ''parameters''. 

Nevertheless, the  results add also to a developing literature on the Hurst exponent value in investment strategies \citep{domino2011use,batten2013structure,horta2014impact,ramos2017introducing,garcia2019different}, especially when there are substantially noticeable changes in market behavior.   It can be concluded that various investor strategies might be deduced or suggested depending on the specific  focus, as suggested in the previous paragraph.

\newpage
\begin{appendices}\label{Predictors}
	\renewcommand{\thesection}{\Alph{section}}
	{ \bf Predictors}
	\section{Description of the macroeconomic variables}\label{sec:Description of the 14 macroeconomic variables}
A brief description of the 14 macroeconomic variables follows.

1. Dividend-price ratio (log), D/P: log of a 12-month
moving sum of dividends paid on the S\&P 500 Index
minus the log of corresponding stock prices (S\&P 500 Index).

2. Dividend yield ratio (log), D/Y: log of a 12-month moving
sum of dividends minus the log of lagged stock
prices.

3. Earnings-price ratio (log), E/P: log of a 12-month
moving sum of earnings on the S\&P 500 Index minus
the log of stock prices.

4. Dividend-payout ratio (log), D/E: log of a 12-month
moving sum of dividends minus the log of a 12-month
moving sum of earnings.

5. Equity risk premium volatility, ERPV: calculated based
on a 12-month moving standard deviation estimator.

6. Book-to-market ratio, B/M: book-to-market value
ratio for the Dow Jones Industrial Average.

7. Net equity expansion ratio, NEER: ratio of a 12-month
moving sum of net equity issues by NYSE-listed stocks
to the total end-of-year market capitalization of New
York Stock Exchange (NYSE) stocks.

8. Treasury bill rate, TBR: interest rate on a secondary market three-month
Treasury bill.

9. Long-term yield, LTY: long-term government bond
yield.

10. Long-term return, LTR: return on long-term
government bonds.

11. Term spread, TMS: long-term yield minus the
Treasury bill rate.

12. Default yield spread, DYS: difference between
Moody’s BAA- and AAA-rated corporate bond yields.

13. Default return spread, DRS: long-term corporate
bond return minus the long-term government bond
return.

14. Inflation, INFL: calculated from the Consumer Price Index ($CPI$) for all urban consumers. To account for the delay in $CPI$ releases, we use its lagged values ($CPI_{t-1}$) in the analysis of this study.

\section{Description of the technical indicators}\label{sec:Description of the technical indicator}
The first is a moving average (MA) rule which generates a buy or sell signal ($S_{i,t}=1$ or $S_{i,t}=0$, respectively) at the end of the month $t$, and is constructed by comparing two moving averages,

\begin{equation}\label{MA rules}
S_{i,t} =
\begin{cases}
1 & \text{if $MA_{s,t}$ $\geq$ $MA_{l,t}$},\\
0 & \text{if $MA_{s,t}$ $<$ $MA_{l,t}$},
\end{cases}       
\end{equation}

where 

\begin{equation}\label{MA}
MA_{j,t} =  \frac{1}{j}\sum_{i=1}^{j-1} P_{t-i}, \;\; \;  \;\;   \; j = s, l;
\end{equation}

$P_{t}$ is the stock Index, $s$ and $l$ denote the length of short and long $MA(s<l)$, respectively. The $MA$ indicator with $MA$ lengths $s$ and $l$ is denoted by $MA(s,l)$. We compute the monthly MA rules with $s = 1,2,3$ and $l=9,12$. We have six $MA$ technical indicators in total which include: MA(1,9), MA(1,12), MA(2,9), MA(2,12), MA(3,9) and MA(3,12).

The second strategy is the momentum rule that generates a buy or sell signal given by

\begin{equation}\label{Momentum rules}
S_{i,t} =
\begin{cases}
1 & \text{if $P_{t}$ $\geq$ $P_{t-m}$},\\
0 & \text{if $P_{t}$ $<$ $P_{t-m}$}.
\end{cases}       
\end{equation}

We denote the momentum indicator that compares $P_{t}$ to $P_{t-m}$ by $MOM(m)$, and compute the monthly signal for $m=9,12$. We have two $MOM$ technical indicators in total which include MOM(9) and MOM(12).

The third strategy is the on-balance volume (OBV), where a buy or sell signal is given by

\begin{equation}\label{OBV rules}
S_{i,t} =
\begin{cases}
1 & \text{if $MA_{s,t}^{OBV}$ $\geq$ $MA_{l,t}^{OBV}$},\\
0 & \text{if $MA_{s,t}^{OBV}$ $<$ $MA_{l,t}^{OBV}$},
\end{cases}       
\end{equation}
where
\begin{equation}\label{MA OBV}
MA_{j,t}^{OBV} =  \frac{1}{j}\sum_{i=1}^{j-1} OBV_{t-i}, \; \;\; \;  \;\;   \; \; j = s, l;
\end{equation}
\begin{equation}\label{OBV}
OBV_{t} =  \sum_{k=1}^{t} VOL_{k}\;D_{k},
\end{equation}

$VOL_{k}$ is a measure of the trading volume during
period $k$, and $D_{k}$ is a binary variable which equals 1 if $P_{k}-P_{k-1} \geq 0$ and $-1$ otherwise. We compute OBV
monthly trading signals for $s=1,2,3$ and $l=9,12$ and denote
the corresponding indicator by $VOL(s,l)$. We obtain the daily volume data from Yahoo Finance and convert it to monthly data.  We have six $VOL$ technical indicators in total which include: VOL(1,9), VOL(1,12), VOL(2,9), VOL(2,12), VOL(3,9) and VOL(3,12).
	
\end{appendices}

\newpage
{\bf Acknowledgements}

  This paper is part of the first author’s Ph.D. thesis. We are grateful for the valuable comments and suggestions from his adviser Dr. G. Charles-Cadogan and two anonymous reviewers which hopefully greatly improved the paper. We also thankfully acknowledge the research support of the University of Leicester, School of Business. Any errors which may remain are our own.

\newpage 
\begin{singlespace}
%	\bibliography{KSUReferences3}

\end{singlespace}
\newpage

\FloatBarrier
% \usepackage{graphicx}
% \use-package{graphicx}
% \usepackage{graphicx}
% \usepackage{graphicx}
% \usepackage{graphicx}
% \usepackage{graphicx}
% \usepackage{graphicx}
\begin{table}[th]
	\centering
	\caption{Summary Statistics}
	\label{tab:summary stat table}
	\resizebox{\textwidth}{!}{%
		\begin{tabular}{lcrrrrrrrrrr}
			\multicolumn{12}{l}{\begin{tabular}[c]{@{}l@{}}Summary of statistical characteristics of the daily log return (in percent) of the S\&P500\\ Index, daily local Hurst exponent, monthly Bullish ratio and monthly Bullish  Index for the\\ sampling period between 1950:12:01 to 2019:12:31.\end{tabular}}                                                              \\ \hline
			Variable             &                      & \multicolumn{1}{c}{Mean} & \multicolumn{1}{c}{} & \multicolumn{1}{c}{Std. Dev.} & \multicolumn{1}{c}{} & \multicolumn{1}{c}{min.} & \multicolumn{1}{c}{} & \multicolumn{1}{c}{max.} & \multicolumn{1}{c}{} & \multicolumn{1}{c}{\begin{tabular}[c]{@{}c@{}}Auto-\\ correlation\end{tabular}} & \multicolumn{1}{c}{} \\ \cline{1-1} \cline{3-3} \cline{5-5} \cline{7-7} \cline{9-9} \cline{11-11}
			Log return           &                      & 0.029                    &                      & 0.965                         &                      & -22.900                  &                      & 10.957                   &                      & 0.025                                                                           &                      \\
			Local Hurst Exponent & \multicolumn{1}{l}{} & 0.557                    &                      & 0.094                         &                      & 0.196                    &                      & 0.786                    &                      & 0.999                                                                           &                      \\
			Bullish Ratio        & \multicolumn{1}{l}{} & 0.396                    &                      & 0.239                         &                      & 0.000                    &                      & 0.882                    &                      & 0.692                                                                           &                      \\
			Bullish Index           &                      & -0.004                   &                      & 0.479                         &                      & -2.848                   &                      & 2.660                    &                      & -0.261                                                                          &                      \\ \hline
			\multicolumn{12}{l}{}                                                                                                                                                                                                                                                                                                                                            
		\end{tabular}%
	}
\end{table}
\newpage
\FloatBarrier
% \usepackage{graphicx}
% \usepackage{graphicx}
% \usepackage{graphicx}
% \usepackage{graphicx}
% \usepackage{graphicx}
% \usepackage{graphicx}
\begin{table}[th]
	\centering
	\caption{Summary Statistics of Equity Risk Premium and Macroeconomic Variables}
	\label{tab:summary stat table macro}
	\resizebox{\textwidth}{!}{%
		\begin{tabular}{llrrrrrrrlr}
			\multicolumn{11}{l}{\begin{tabular}[c]{@{}l@{}}Summary of statistical characteristics of the log equity risk premium (in percent) and 14 macroeconomic\\ variables for the sampling period between 1950:12:01 to 2019:12:31. LTR, DRS and INFL  measured in percent \\while  TBR, LTY,   TMS, and DYS are measured in annual percent.\end{tabular}}                                       \\ \hline
			(i) Variable                                                           & \multicolumn{1}{c}{} & \multicolumn{1}{c}{Mean} & \multicolumn{1}{c}{} & \multicolumn{1}{c}{Std. Dev.} & \multicolumn{1}{c}{} & \multicolumn{1}{c}{Min} & \multicolumn{1}{c}{} & \multicolumn{1}{c}{Max} &  & \multicolumn{1}{l}{\begin{tabular}[c]{@{}l@{}}Auto-\\ correlation\end{tabular}} \\ \cline{1-1} \cline{3-3} \cline{5-5} \cline{7-7} \cline{9-9} \cline{11-11} 
			\begin{tabular}[c]{@{}l@{}}\;\;\;\;\;\;Log equity   risk premium\end{tabular} & \multicolumn{1}{c}{} & 0.539                    &                      & 4.152                         &                      & -24.840                 &                      & 14.870                  &  & 0.045                                                                           \\
			\hline
			(1) Dividend-price ratio (D/P)                          & \multicolumn{1}{c}{} & -3.540                   &                      & 0.415                         &                      & -4.524                  &                      & -2.598                  &  & 0.995                                                                           \\
			(2) Dividend yield (D/Y)                                               &                      & -3.533                   &                      & 0.416                         &                      & -4.531                  &                      & -2.586                  &  & 0.995                                                                           \\
			(3) Earnings-price ratio (E/P)                                         &                      & -2.806                   &                      & 0.420                         &                      & -4.836                  &                      & -1.899                  &  & 0.990                                                                           \\
			(4) Dividend-payout ratio (D/E)                                        &                      & -0.734                   &                      & 0.290                         &                      & -1.244                  &                      & 1.380                   &  & 0.987                                                                           \\
			(5) Equity risk premium volatility (ERPV)                              &                      & 0.141                    &                      & 0.048                         &                      & 0.053                   &                      & 0.316                   &  & 0.961                                                                           \\
			(6) Book-to-market ratio (B/M)                                         &                      & 0.509                    &                      & 0.247                         &                      & 0.121                   &                      & 1.207                   &  & 0.994                                                                           \\
			(7) Net equity expansion ratio (NEER)                                  &                      & 0.012                    &                      & 0.020                         &                      & -0.058                  &                      & 0.051                   &  & 0.982                                                                           \\
			(8) Treasury bill rate (TBR)                                           &                      & 4.213                    &                      & 3.071                         &                      & 0.010                   &                      & 16.300                  &  & 0.991                                                                           \\
			(9) Long-term yield (LTY)                                              &                      & 5.892                    &                      & 2.781                         &                      & 1.630                   &                      & 14.820                  &  & 0.995                                                                           \\
			(10) Long-term return (LTR)                                             &                      & 0.530                    &                      & 2.762                         &                      & -11.240                 &                      & 15.230                  &  & 0.042                                                                           \\
			(11) Term spread (TMS)                                                  &                      & 1.679                    &                      & 1.387                         &                      & -3.650                  &                      & 4.550                   &  & 0.958                                                                           \\
			(12) Default yield spread (DYS)                                         &                      & 0.960                    &                      & 0.434                         &                      & 0.320                   &                      & 3.380                   &  & 0.971                                                                           \\
			(13) Default return spread (DRS)                                        &                      & 0.030                    &                      & 1.397                         &                      & -9.750                  &                      & 7.370                   &  & -0.077                                                                          \\
			(14) Inflation (INFL)                                                   &                      & 0.284                    &                      & 0.358                         &                      & -1.915                  &                      & 1.806                   &  & 0.548                                                                           \\ \hline
		\end{tabular}%
	}
\end{table}
%\hspace*{-1.5cm}\resizebox{26cm}{!} 	\thispagestyle{empty}

\newpage

% \usepackage{multirow}
% \usepackage{graphicx}
% \usepackage{lscape}
% \usepackage{multirow}  		\thispagestyle{empty}
% \usepackage{graphicx}  \hspace*{-1.5cm}\resizebox{26cm}{!}

% \usepackage{multirow}
% \usepackage{graphicx}
% \usepackage{lscape}
% \usepackage{multirow}
% \usepackage{graphicx}
% \usepackage{lscape}
\begin{landscape}
	\begin{table}[]
		\centering
		\caption{In-sample Forecasting Results }
		\label{tab:In-sampleResults}
		\thispagestyle{empty}
		\hspace*{-1.5cm}\resizebox{26cm}{!}{%
			\begin{tabular}{llllrrlrrlrrlllllrrlrrlrr}
				\multicolumn{25}{l}{Panel A reports the in-sample estimation results of the following bivariate predictive regression for the sampling period from 1951:01 to 2019:12.}                                                                                                                                                                                                                                                                                                                                                                                                                                                                                                                                                                                                                                                                                                                                                                                                                                                                     \\
				\multicolumn{25}{c}{$r_{t+1}=\alpha^{(i)}+\beta^{(i)} \psi^{(i)}_{t}+\varepsilon^{(i)}_{t+1}$,}                                                                                                                                                                                                                                                                                                                                                                                                                                                                                                                                                                                                                                                                                                                                                                                                                                                                                                                                                           \\
				\multicolumn{25}{l}{\begin{tabular}[c]{@{}l@{}}where $r_{t+1}$ is the monthly equity risk premium (in percent) and $\psi_{i,t}$ is one of the 14 macroeconomic variables or one of the 14 technical indicators. Panel B and C report estimation results for a predictive regression\\ based on principal components as predictor,\end{tabular}}                                                                                                                                                                                                                                                                                                                                                                                                                                                                                                                                                                                                                                                                                            \\
				\multicolumn{25}{c}{$r_{t+1}=\alpha^{(j)}+\sum_{k=1}^{K}\gamma_{k} \, \hat{F}^{(j)}_{k,t}+\varepsilon^{(j)}_{t+1}$.}                                                                                                                                                                                                                                                                                                                                                                                                                                                                                                                                                                                                                                                                                                                                                                                                                                                                                                                        \\
				\multicolumn{25}{l}{\begin{tabular}[c]{@{}l@{}}where $\hat{F}_{k,t}^{j}$ is the first $K$ principal components estimated from the 14 macroeconomic variables $(j=ECON)$, 14 technical indicators $(j=TECH)$, or all of the 14 macroeconomic variables and 14 technical indicators\\ taken together $(j=ALL)$. We present the estimates of regression slope coefficients, Newey and West \citep{newey1986simple} t-statistics in parentheses, and $R^{2}$ statistics (in percent) for the entire sampling period. The $R^{2}$ statistics \\ (in percent) is reported for the entire sampling period, business-cycle expansions ($exp$) and recessions ($rec$). The $R^{2}_{BU+}$ ($R^{2}_{BU-}$) is computed for the Bullish Index peak (trough) periods as described in the text. The\\ corresponding stable periods are denoted by $stable^{+}$ ($stable^{-}$). The *, ** and *** indicate significance at the 10\%, 5\% and 1\% levels, respectively, based on one-sided (upper-tail) wild bootstrapped p-values.\end{tabular}} \\ \hline
				\multicolumn{12}{c}{Macroeconomic variables}                                                                                                                                                                                                                                                                                                                                                                                                                                                                                     &    & \multicolumn{12}{c}{Technical indicators}                                                                                                                                                                                                                                                                                                                                                                                                                                                         \\ \cline{1-12} \cline{14-25} 
				\multicolumn{25}{l}{}                                                                                                                                                                                                                                                                                                                                                                                                                                                                                                                                                                                                                                                                                                                                                                                                                                                                                                                                                                                                                     \\
				\multicolumn{3}{c}{Entire period}                                                                   &                         & \multicolumn{2}{c}{Recession dummy}                                                                           &                         & \multicolumn{2}{c}{Bullish Index peak}                                                                                     &     & \multicolumn{2}{c}{Bullish Index trough}                                                                                  &    & \multicolumn{3}{c}{Entire period}                                                                &                        & \multicolumn{2}{c}{Recession dummy}                                                                         &    & \multicolumn{2}{c}{Bullish Index peak}                                                                                   &    & \multicolumn{2}{c}{Bullish Index trough}                                                                                 \\ \cline{1-3} \cline{5-6} \cline{8-9} \cline{11-12} \cline{14-16} \cline{18-19} \cline{21-22} \cline{24-25} 
				\multirow{2}{*}{Predictor}    & \multirow{2}{*}{Slope coefficient}    & \multirow{2}{*}{$R^{2}$}    &                         & \multicolumn{1}{l}{\multirow{2}{*}{$R^{2}_{exp}$}}    & \multicolumn{1}{c}{\multirow{2}{*}{$R^{2}_{rec}$}}    &                         & \multicolumn{1}{l}{\multirow{2}{*}{$R^{2}_{stable^{+}}$}}    & \multicolumn{1}{l}{\multirow{2}{*}{$R^{2}_{BU^{+}}$}}    &     & \multicolumn{1}{l}{\multirow{2}{*}{$R^{2}_{stable^{-}}$}}    & \multicolumn{1}{l}{\multirow{2}{*}{$R^{2}_{BU^{-}}$}}   &    & \multirow{2}{*}{Predictor}   & \multirow{2}{*}{Slope coefficient}   & \multirow{2}{*}{$R^{2}$}   &                        & \multicolumn{1}{l}{\multirow{2}{*}{$R^{2}_{exp}$}}   & \multicolumn{1}{c}{\multirow{2}{*}{ $R^{2}_{rec}$}}   &    & \multicolumn{1}{l}{\multirow{2}{*}{$R^{2}_{stable^{+}}$}}   & \multicolumn{1}{l}{\multirow{2}{*}{$R^{2}_{BU^{+}}$}}   &    & \multicolumn{1}{l}{\multirow{2}{*}{$R^{2}_{stable^{-}}$}}   & \multicolumn{1}{l}{\multirow{2}{*}{$R^{2}_{BU^{-}}$}}   \\
				&                                       &                             &                         & \multicolumn{1}{l}{}                                  & \multicolumn{1}{l}{}                                  &                         & \multicolumn{1}{l}{}                                         & \multicolumn{1}{l}{}                                     &     & \multicolumn{1}{l}{}                                         & \multicolumn{1}{l}{}                                    &    &                              &                                      &                            &                        & \multicolumn{1}{l}{}                                 & \multicolumn{1}{l}{}                                 &    & \multicolumn{1}{l}{}                                        & \multicolumn{1}{l}{}                                    &    & \multicolumn{1}{l}{}                                        & \multicolumn{1}{l}{}                                    \\ \hline
				\multicolumn{1}{c}{}          &                                       &                             &                         & \multicolumn{1}{l}{}                                  & \multicolumn{1}{l}{}                                  &                         & \multicolumn{1}{l}{}                                         & \multicolumn{1}{l}{}                                     &     & \multicolumn{1}{l}{}                                         & \multicolumn{1}{l}{}                                    &    &                              &                                      &                            &                        & \multicolumn{1}{l}{}                                 & \multicolumn{1}{l}{}                                 &    & \multicolumn{1}{l}{}                                        & \multicolumn{1}{l}{}                                    &    & \multicolumn{1}{l}{}                                        & \multicolumn{1}{l}{}                                    \\
				\multicolumn{25}{c}{Panel A: Bivariate predictive regressions}                                                                                                                                                                                                                                                                                                                                                                                                                                                                                                                                                                                                                                                                                                                                                                                                                                                                                                                                                                            \\
				D/P                           & 0.56(1.55)                            & 0.32                        &                         & 0.22                                                  & 0.55                                                  &                         & 0.32                                                         & 0.19                                                     &     & 0.37                                                         & 0.11                                                    &    & MA(1,9)                      & 0.58(1.60)**                         & 0.40                       &                        & -0.44                                                & 2.52                                                 &    & 0.47                                                        & 0.25                                                    &    & 0.22                                                        & 1.18                                                    \\
				D/Y                           & 0.60(1.67)**                          & 0.36                        &                         & 0.15                                                  & 0.90                                                  &                         & 0.39                                                         & 0.16                                                     &     & 0.42                                                         & 0.12                                                    &    & MA(1,12)                     & 0.75(1.99)**                         & 0.66                       &                        & -0.34                                                & 3.16                                                 &    & 0.31                                                        & 1.99                                                    &    & 0.31                                                        & 2.10                                                    \\
				E/P                           & 0.38(0.89)                            & 0.15                        &                         & 0.18                                                  & 0.06                                                  &                         & 0.17                                                         & 0.05                                                     &     & 0.15                                                         & 0.12                                                    &    & MA(2,9)                      & 0.67(1.87)**                         & 0.54                       &                        & -0.29                                                & 2.63                                                 &    & 0.31                                                        & 1.47                                                    &    & 0.30                                                        & 1.57                                                    \\
				D/E                           & 0.36(0.59)                            & 0.06                        &                         & 0.01                                                  & 0.20                                                  &                         & 0.02                                                         & 0.15                                                     &     & 0.08                                                         & 0.00                                                    &    & MA(2,12)                     & 0.90(2.39)***                        & 0.93                       &                        & -0.17                                                & 3.71                                                 &    & 0.57                                                        & 2.39                                                    &    & 0.51                                                        & 2.71                                                    \\
				ERPV                          & 6.30(2.18)**                          & 0.54                        &                         & 0.40                                                  & 0.89                                                  &                         & 0.47                                                         & 0.63                                                     &     & 0.89                                                         & -0.95                                                   &    & MA(3,9)                      & 0.68(1.89)**                         & 0.55                       &                        & -0.12                                                & 2.24                                                 &    & 0.35                                                        & 1.38                                                    &    & 0.27                                                        & 1.76                                                    \\
				B/M                           & 0.26(0.40)                            & 0.02                        & \multicolumn{1}{r}{}    & -0.01                                                 & 0.11                                                  &                         & 0.01                                                         & 0.04                                                     &     & 0.07                                                         & -0.17                                                   &    & MA(3,12)                     & 0.43(1.17)                           & 0.22                       &                        & -0.21                                                & 1.30                                                 &    & 0.11                                                        & 0.67                                                    &    & -0.15                                                       & 1.76                                                    \\
				NEER                          & 4.04(0.48)                            & 0.04                        &                         & 0.24                                                  & -0.47                                                 &                         & -0.06                                                        & 0.43                                                     &     & -0.02                                                        & 0.26                                                    &    & MOM(9)                       & 0.45(1.21)*                          & 0.24                       &                        & -0.18                                                & 1.28                                                 &    & 0.13                                                        & 0.66                                                    &    & 0.01                                                        & 1.21                                                    \\
				TBR                           & 0.12(2.39)***                         & 0.76                        &                         & 0.57                                                  & 1.26                                                  &                         & 0.76                                                         & 0.75                                                     &     & 0.57                                                         & 1.58                                                    &    & MOM(12)                      & 0.49(1.29)*                          & 0.27                       &                        & -0.39                                                & 1.93                                                 &    & 0.15                                                        & 0.76                                                    &    & 0.00                                                        & 1.39                                                    \\
				LTY                           & 0.10(1.80)**                          & 0.41                        &                         & 0.39                                                  & 0.48                                                  &                         & 0.64                                                         & -0.37                                                    &     & 0.30                                                         & 0.89                                                    &    & VOL(1,9)                     & 0.63(1.80)**                         & 0.49                       &                        & -0.50                                                & 2.98                                                 &    & 0.25                                                        & 1.44                                                    &    & 0.28                                                        & 1.36                                                    \\
				LTR                           & 0.14(2.38)***                         & 0.91                        & \multicolumn{1}{r}{}    & -0.13                                                 & 3.52                                                  &                         & 0.81                                                         & 1.17                                                     &     & 0.89                                                         & 1.03                                                    &    & VOL(1,12)                    & 0.79(2.16)**                         & 0.74                       &                        & -0.31                                                & 3.39                                                 &    & 0.74                                                        & 0.85                                                    &    & 0.37                                                        & 2.33                                                    \\
				TMS                           & 0.19(1.79)**                          & 0.42                        &                         & 0.02                                                  & 1.43                                                  &                         & -0.04                                                        & 2.02                                                     &     & 0.26                                                         & 1.09                                                    &    & VOL(2,9)                     & 0.60(1.75)**                         & 0.46                       &                        & -0.25                                                & 2.22                                                 &    & 0.24                                                        & 1.32                                                    &    & 0.09                                                        & 1.99                                                    \\
				DYS                           & 0.14(0.34)                            & 0.02                        &                         & 0.04                                                  & -0.02                                                 &                         & -0.10                                                        & 0.48                                                     &     & 0.06                                                         & -0.13                                                   &    & VOL(2,12)                    & 0.64(1.76)**                         & 0.49                       &                        & -0.17                                                & 2.16                                                 &    & 0.37                                                        & 1.02                                                    &    & 0.12                                                        & 2.07                                                    \\
				DRS                           & 0.14(0.89)                            & 0.23                        &                         & 0.03                                                  & 0.72                                                  &                         & 0.30                                                         & -0.07                                                    &     & 0.27                                                         & 0.03                                                    &    & VOL(3,9)                     & 0.41(1.17)                           & 0.21                       &                        & -0.19                                                & 1.22                                                 &    & 0.20                                                        & 0.30                                                    &    & -0.06                                                       & 1.34                                                    \\
				INFL                          & 0.29(0.58)                            & 0.06                        &                         & 0.22                                                   & -0.35                                                 &                         & -0.02                                                        & 0.30                                                     &     & -0.03                                                        & 0.46                                                    &    & VOL(3,12)                    & 0.75(2.12)**                         & 0.68                       &                        & 0.03                                                 & 2.31                                                 &    & 0.80                                                        & 0.23                                                    &    & 0.41                                                        & 1.82                                                    \\
				\multicolumn{1}{c}{}          &                                       &                             &                         & \multicolumn{1}{l}{}                                  & \multicolumn{1}{l}{}                                  &                         & \multicolumn{1}{l}{}                                         & \multicolumn{1}{l}{}                                     &     & \multicolumn{1}{l}{}                                         & \multicolumn{1}{l}{}                                    &    &                              &                                      &                            &                        & \multicolumn{1}{l}{}                                 & \multicolumn{1}{l}{}                                 &    & \multicolumn{1}{l}{}                                        & \multicolumn{1}{l}{}                                    &    & \multicolumn{1}{l}{}                                        & \multicolumn{1}{l}{}                                    \\
				\multicolumn{25}{c}{Panel B: Principal component predictive regressions}                                                                                                                                                                                                                                                                                                                                                                                                                                                                                                                                                                                                                                                                                                                                                                                                                                                                                                                                                                  \\
				$\hat{F}_{1}^{ECON}$          & 0.01(0.06)                            & 0.73                        &                         & 0.51                                                  & 1.29                                                  &                         & 0.49                                                         & 1.35                                                     &     & 0.74                                                         & 0.71                                                    &    & $\hat{F}_{1}^{TECH}$         & 0.11(1.95)**                         & 0.66                       &                        & -0.32                                                & 3.12                                                 &    & 0.48                                                        & 1.42                                                    &    & 0.23                                                        & 2.49                                                    \\
				\multicolumn{1}{c}{}          & \multicolumn{1}{c}{}                  & \multicolumn{1}{c}{}        & \multicolumn{1}{c}{}    & \multicolumn{1}{c}{}                                  & \multicolumn{1}{c}{}                                  & \multicolumn{1}{c}{}    & \multicolumn{1}{l}{}                                         & \multicolumn{1}{l}{}                                     &     & \multicolumn{1}{l}{}                                         & \multicolumn{1}{l}{}                                    &    & \multicolumn{1}{c}{}         & \multicolumn{1}{c}{}                 & \multicolumn{1}{c}{}       & \multicolumn{1}{c}{}   & \multicolumn{1}{c}{}                                 & \multicolumn{1}{c}{}                                 &    & \multicolumn{1}{l}{}                                        & \multicolumn{1}{l}{}                                    &    & \multicolumn{1}{l}{}                                        & \multicolumn{1}{l}{}                                    \\
				\multicolumn{25}{c}{Panel C: Principal component predictive regression, all predictors taken together}                                                                                                                                                                                                                                                                                                                                                                                                                                                                                                                                                                                                                                                                                                                                                                                                                                                                                                                                \\
				$\hat{F}_{1}^{ALL}$           & 0.10(1.83)**                          & 1.41                        &                         & -0.04                                                 & 5.04                                                  &                         & 1.01                                                         & 2.72                                                     &     & 1.00                                                         & 3.13                                                    &    &                              &                                      &                            &                        & \multicolumn{1}{l}{}                                 & \multicolumn{1}{l}{}                                 &    & \multicolumn{1}{l}{}                                        & \multicolumn{1}{l}{}                                    &    & \multicolumn{1}{l}{}                                        & \multicolumn{1}{l}{}                                    \\ \hline
			\end{tabular}%
		}
	\end{table}
\end{landscape}

\begin{landscape}
	\begin{table}[]
		\centering
		\caption{Out-of-sample Forecasting Results }
		\label{tab:oosResults}
		\thispagestyle{empty}
		\hspace*{-1.5cm}\resizebox{26cm}{!}{%
			\begin{tabular}{llrrrrrrrrrrrllrrrrrrrrrr}
				\multicolumn{25}{l}{Table \ref{tab:oosResults} reports the out-of-sample forecasting results from 1966:01 to 2019:12. Panel A reports the following bivariate predictive regression forecasts for each of the predictors,}                                                                                                                                                                                                                                                                                                                                                                                                                                                                                                                                                                                                                                                                                                                                                                                                                                                                                                                                                                                                                                                                                                                                                                 \\
				\multicolumn{25}{c}{$\hat{r}_{t+1}=\hat{\alpha}^{(i)}_{t}+\hat{\beta}^{(i)}_{t} \psi^{(i)}_{t}$,}                                                                                                                                                                                                                                                                                                                                                                                                                                                                                                                                                                                                                                                                                                                                                                                                                                                                                                                                                                                                                                                                                                                                                                                                                                                                                      \\
				\multicolumn{25}{l}{\begin{tabular}[c]{@{}l@{}}where $\hat{r}_{t+1}$ is the log equity premium (in percent) and $\psi^{(i)}_{t}$ is one of the 14 macroeconomic variables or one of the 14 technical indicators. Panel B and C report the PC-ECON,\\ PC-TECH and PC-ALL forecasts given by\end{tabular}}                                                                                                                                                                                                                                                                                                                                                                                                                                                                                                                                                                                                                                                                                                                                                                                                                                                                                                                                                                                                                                                                               \\
				\multicolumn{25}{c}{$\hat{r}_{t+1}=\hat{\alpha}^{(j)}_{t}+\sum_{k=1}^{K}\hat{\gamma}_{k} \hat{F}_{1:k,k,t}^{(j)}$,}                                                                                                                                                                                                                                                                                                                                                                                                                                                                                                                                                                                                                                                                                                                                                                                                                                                                                                                                                                                                                                                                                                                                                                                                                                                                \\
				\multicolumn{25}{l}{\begin{tabular}[c]{@{}l@{}}where $\hat{F}_{k,t}^{j}$ is the $k$th principal component extracted from the 14 macroeconomic variables $(j=ECON)$, 14 technical indicators $(j=TECH)$, or the 14 macroeconomic variables and\\ 14 technical indicators taken together $(j=ALL)$. We select $K$ via the adjusted $R^{2}$ based on data through $t$. We present the adjusted mean squared forecast error (MSFE-adjusted) of \\ Clark and West \citep{clark2007approximately}. The $R^{2}_{OS}$ statistics (in percent) is reported for the entire sampling period, business-cycle expansions (exp) and recessions (rec). The $R^{2}_{OS,BU+}$ ($R^{2}_{OS,BU-}$) is\\ computed for the Bullish Index peak (trough) periods as described in the text. The corresponding stable periods are denoted by $stable^{+}$ ($stable^{-}$). The *, ** and *** indicate \\ significance at the 10\%, 5\% and 1\% levels,respectively.\end{tabular}}                                                                                                                                                                                                                                                                                                                                                                                                                                       \\ \hline
				\multicolumn{12}{c}{Macroeconomic variables}                                                                                                                                                                                                                                                                                                                                                                                                                                                                                                                                                                                                                                                 &  & \multicolumn{12}{c}{Technical indicators}                                                                                                                                                                                                                                                                                                                                                                                                                                                                                                                                                                                                                                \\ \cline{1-12} \cline{14-25} 
				\multicolumn{25}{l}{}                                                                                                                                                                                                                                                                                                                                                                                                                                                                                                                                                                                                                                                                                                                                                                                                                                                                                                                                                                                                                                                                                                                                                                                                                                                                                                                                                      \\
				\multicolumn{3}{c}{Entire period}                                                                                                      &                      & \multicolumn{2}{c}{\begin{tabular}[c]{@{}c@{}}Recession\\ dummy\end{tabular}}                                                                                  &  & \multicolumn{2}{c}{\begin{tabular}[c]{@{}c@{}}Bullish Index \\ peak\end{tabular}}                                                                                            &  & \multicolumn{2}{c}{\begin{tabular}[c]{@{}c@{}}Bullish Index\\  trough\end{tabular}}                                                                                          &  & \multicolumn{3}{c}{Entire period}                                                                                                      &  & \multicolumn{2}{c}{\begin{tabular}[c]{@{}c@{}}Recession \\ dummy\end{tabular}}                                                                                 &  & \multicolumn{2}{c}{\begin{tabular}[c]{@{}c@{}}Bullish Index \\ peak\end{tabular}}                                                                                            &  & \multicolumn{2}{c}{\begin{tabular}[c]{@{}c@{}}Bullish Index \\ trough\end{tabular}}                                                                                          \\ \cline{1-3} \cline{5-6} \cline{8-9} \cline{11-12} \cline{14-16} \cline{18-19} \cline{21-22} \cline{24-25} 
				\multirow{2}{*}{Predictor} & \multirow{2}{*}{\begin{tabular}[c]{@{}l@{}}MSFE-\\ adjusted\end{tabular}} & \multirow{2}{*}{$R^{2}_{OS}$} &                      & \multirow{2}{*}{\begin{tabular}[c]{@{}l@{}}$R^{2}_{OS}$,\\ $exp$\end{tabular}} & \multirow{2}{*}{\begin{tabular}[c]{@{}l@{}}$R^{2}_{OS}$\\ $rec$\end{tabular}} &  & \multirow{2}{*}{\begin{tabular}[c]{@{}l@{}}$R^{2}_{OS}$,\\ $stable^{+}$\end{tabular}} & \multirow{2}{*}{\begin{tabular}[c]{@{}l@{}}$R^{2}_{OS}$,\\ $BU^{+}$\end{tabular}} &  & \multirow{2}{*}{\begin{tabular}[c]{@{}l@{}}$R^{2}_{OS}$,\\ $stable^{-}$\end{tabular}} & \multirow{2}{*}{\begin{tabular}[c]{@{}l@{}}$R^{2}_{OS}$,\\ $BU^{-}$\end{tabular}} &  & \multirow{2}{*}{Predictor} & \multirow{2}{*}{\begin{tabular}[c]{@{}l@{}}MSFE-\\ adjusted\end{tabular}} & \multirow{2}{*}{$R^{2}_{OS}$} &  & \multirow{2}{*}{\begin{tabular}[c]{@{}l@{}}$R^{2}_{OS}$,\\ $exp$\end{tabular}} & \multirow{2}{*}{\begin{tabular}[c]{@{}l@{}}$R^{2}_{OS}$\\ $rec$\end{tabular}} &  & \multirow{2}{*}{\begin{tabular}[c]{@{}l@{}}$R^{2}_{OS}$,\\ $stable^{+}$\end{tabular}} & \multirow{2}{*}{\begin{tabular}[c]{@{}l@{}}$R^{2}_{OS}$,\\ $BU^{+}$\end{tabular}} &  & \multirow{2}{*}{\begin{tabular}[c]{@{}l@{}}$R^{2}_{OS}$,\\ $stable^{-}$\end{tabular}} & \multirow{2}{*}{\begin{tabular}[c]{@{}l@{}}$R^{2}_{OS}$,\\ $BU^{-}$\end{tabular}} \\
				&                                                                           &                               &                      &                                                                                &                                                                               &  &                                                                                       &                                                                                   &  &                                                                                       &                                                                                   &  &                            &                                                                           &                               &  &                                                                                &                                                                               &  &                                                                                       &                                                                                   &  &                                                                                       &                                                                                   \\ \hline
				\multicolumn{25}{c}{}                                                                                                                                                                                                                                                                                                                                                                                                                                                                                                                                                                                                                                                                                                                                                                                                                                                                                                                                                                                                                                                                                                                                                                                                                                                                                                                                                      \\
				\multicolumn{25}{c}{Panel A: Bivariate predictive regression forecasts}                                                                                                                                                                                                                                                                                                                                                                                                                                                                                                                                                                                                                                                                                                                                                                                                                                                                                                                                                                                                                                                                                                                                                                                                                                                                                                    \\
				D/P                        & 1.01                                                                      & -0.30                         &                      & -1.09                                                                          & 1.55                                                                          &  & -0.72                                                                                 & 5.42                                                                              &  & -0.58                                                                                 & 1.88                                                                              &  & MA(1,9)                    & 0.81                                                                      & 0.09                          &  & -0.90                                                                          & 2.39                                                                          &  & -0.01                                                                                 & 1.44                                                                              &  & 0.01                                                                                  & 0.68                                                                              \\
				D/Y                        & 1.12                                                                      & -0.27                         &                      & -1.41                                                                          & 2.39                                                                          &  & -0.68                                                                                 & 5.26                                                                              &  & -0.53                                                                                 & 1.72                                                                              &  & MA(1,12)                   & 1.32*                                                                     & 0.39                          &  & -0.87                                                                          & 3.30                                                                          &  & 0.35                                                                                  & 0.92                                                                              &  & 0.19                                                                                  & 1.89                                                                              \\
				E/P                        & \hspace*{-0.1cm}-0.03                                                                     & -0.59                         &                      & -0.34                                                                          & -1.18                                                                         &  & -0.67                                                                                 & 0.46                                                                              &  & -0.79                                                                                 & 0.93                                                                              &  & MA(2,9)                    & 1.16                                                                      & 0.27                          &  & -0.70                                                                          & 2.54                                                                          &  & 0.23                                                                                  & 0.81                                                                              &  & 0.14                                                                                  & 1.25                                                                              \\
				D/E                        & 0.66                                                                      & -0.84                         &                      & -1.58                                                                          & 0.88                                                                          &  & -0.77                                                                                 & -1.78                                                                             &  & -0.56                                                                                 & -3.02                                                                             &  & MA(2,12)                   & 1.68**                                                                    & 0.67                          &  & -0.56                                                                          & 3.53                                                                          &  & 0.71                                                                                  & 0.12                                                                              &  & 0.49                                                                                  & 2.09                                                                              \\
				ERPV                       & 1.44*                                                                     & -0.01                         &                      & -0.25                                                                          & 0.53                                                                          &  & -0.06                                                                                 & 0.63                                                                              &  & 0.35                                                                                  & -2.83                                                                             &  & MA(3,9)                    & 1.30*                                                                     & 0.25                          &  & -0.90                                                                          & 2.94                                                                          &  & 0.26                                                                                  & 0.21                                                                              &  & 0.07                                                                                  & 1.64                                                                              \\
				B/M                        & \hspace*{-0.1cm}-1.42                                                                     & -1.23                         & \multicolumn{1}{r}{} & -0.33                                                                          & -3.31                                                                         &  & -1.05                                                                                 & -3.61                                                                             &  & -1.03                                                                                 & -2.71                                                                             &  & MA(3,12)                   & 0.44                                                                      & -0.06                         &  & -0.59                                                                          & 1.19                                                                          &  & 0.02                                                                                  & -1.06                                                                             &  & -0.26                                                                                 & 1.53                                                                              \\
				NEER                       & 0.36                                                                      & -0.86                         &                      & -0.12                                                                          & -2.60                                                                         &  & -1.21                                                                                 & 3.78                                                                              &  & -1.07                                                                                 & 0.75                                                                              &  & MOM(9)                     & 0.42                                                                      & -0.02                         &  & -0.60                                                                          & 1.34                                                                          &  & 0.05                                                                                  & -0.89                                                                             &  & -0.25                                                                                 & 1.80                                                                              \\
				TBR                        & 2.22***                                                                   & -0.77                         &                      & -1.71                                                                          & 1.42                                                                          &  & -0.04                                                                                 & -10.60                                                                            &  & -0.29                                                                                 & -4.41                                                                             &  & MOM(12)                    & 0.51                                                                      & 0.04                          &  & -0.54                                                                          & 1.39                                                                          &  & 0.08                                                                                  & -0.46                                                                             &  & -0.07                                                                                 & 0.86                                                                              \\
				LTY                        & 1.71**                                                                    & -0.65                         &                      & -1.35                                                                          & 0.98                                                                          &  & 0.01                                                                                  & -9.57                                                                             &  & -0.40                                                                                 & -2.59                                                                             &  & VOL(1,9)                   & 1.20                                                                      & 0.31                          &  & -0.67                                                                          & 2.61                                                                          &  & 0.41                                                                                  & -0.96                                                                             &  & 0.14                                                                                  & 1.63                                                                              \\
				LTR                        & 2.23***                                                                   & 0.49                          & \multicolumn{1}{r}{} & -1.41                                                                          & 4.90                                                                          &  & 0.00                                                                                  & 7.12                                                                              &  & 0.03                                                                                  & 4.05                                                                              &  & VOL(1,12)                  & 1.60*                                                                     & 0.58                          &  & -0.43                                                                          & 2.94                                                                          &  & 0.85                                                                                  & -3.07                                                                             &  & 0.36                                                                                  & 2.31                                                                              \\
				TMS                        & 2.10**                                                                    & -0.91                         &                      & -3.04                                                                          & 4.06                                                                          &  & -1.37                                                                                 & 5.34                                                                              &  & -0.94                                                                                 & -0.70                                                                             &  & VOL(2,9)                   & 1.03                                                                      & 0.22                          &  & -0.25                                                                          & 1.31                                                                          &  & 0.28                                                                                  & -0.63                                                                             &  & 0.02                                                                                  & 1.73                                                                              \\
				DYS                        & \hspace*{-0.1cm}-0.36                                                                     & -0.61                         &                      & -0.51                                                                          & -0.83                                                                         &  & -0.78                                                                                 & 1.73                                                                              &  & -0.48                                                                                 & -1.54                                                                             &  & VOL(2,12)                  & 0.85                                                                      & 0.15                          &  & -0.04                                                                          & 0.58                                                                          &  & 0.33                                                                                  & -2.31                                                                             &  & 0.06                                                                                  & 0.81                                                                              \\
				DRS                        & 0.05                                                                      & -0.46                         &                      & 0.22                                                                           & -2.06                                                                         &  & -0.69                                                                                 & 2.55                                                                              &  & -1.11                                                                                 & 4.50                                                                              &  & VOL(3,9)                   & 0.27                                                                      & -0.06                         &  & -0.38                                                                          & 0.66                                                                          &  & 0.07                                                                                  & -1.93                                                                             &  & -0.12                                                                                  & 0.37                                                                              \\
				INFL                       & 0.39                                                                      & -0.33                         &                      & -0.10                                                                          & -0.88                                                                         &  & -0.25                                                                                 & -1.44                                                                             &  & 0.05                                                                                  & -3.26                                                                             &  & VOL(3,12)                  & 1.52*                                                                     & 0.51                          &  & -0.09                                                                          & 1.91                                                                          &  & 0.75                                                                                  & -2.68                                                                             &  & 0.42                                                                                  & 1.18                                                                              \\
				\multicolumn{25}{c}{}                                                                                                                                                                                                                                                                                                                                                                                                                                                                                                                                                                                                                                                                                                                                                                                                                                                                                                                                                                                                                                                                                                                                                                                                                                                                                                                                                      \\
				\multicolumn{25}{c}{Panel B: Principal component predictive regression forecasts}                                                                                                                                                                                                                                                                                                                                                                                                                                                                                                                                                                                                                                                                                                                                                                                                                                                                                                                                                                                                                                                                                                                                                                                                                                                                                          \\
				PC-ECON                    & 2.48***                                                                   & -1.11                         &                      & -4.03                                                                          & 5.68                                                                          &  & -1.91                                                                                 & 9.64                                                                              &  & -0.84                                                                                 & -3.24                                                                             &  & PC-TECH                    & 1.31*                                                                     & 0.44                          &  & -0.58                                                                          & 2.81                                                                          &  & 0.55                                                                                  & -1.00                                                                             &  & 0.23                                                                                  & 2.10                                                                              \\
				\multicolumn{25}{c}{}                                                                                                                                                                                                                                                                                                                                                                                                                                                                                                                                                                                                                                                                                                                                                                                                                                                                                                                                                                                                                                                                                                                                                                                                                                                                                                                                                      \\
				\multicolumn{25}{c}{Panel C: Principal component predictive regression forecasts, all predictors taken together}                                                                                                                                                                                                                                                                                                                                                                                                                                                                                                                                                                                                                                                                                                                                                                                                                                                                                                                                                                                                                                                                                                                                                                                                                                                                      \\
				PC-ALL                     & 3.07***                                                                   & 1.14                          &                      & -3.20                                                                          & 11.26                                                                         &  & 0.35                                                                                  & 11.89                                                                             &  & 0.59                                                                                  & 5.42                                                                              &  &                            &                                                                           &                               &  &                                                                                &                                                                               &  &                                                                                       &                                                                                   &  &                                                                                       &                                                                                   \\ \hline
			\end{tabular}%
		}
	\end{table}
\end{landscape}

%	\hspace*{-1.5cm}\resizebox{26cm} 	\thispagestyle{empty}
\newpage

% \usepackage{multirow}
% \usepackage{graphicx}
% \usepackage{lscape}
\begin{landscape}
	\begin{table}[]
		\centering
		\caption{Asset Allocation Performance }
		\label{tab:assetallocation}
			\thispagestyle{empty}
	\hspace*{-1.5cm}\resizebox{26cm}{!}{%
			\begin{tabular}{lrrrrrrrrrrrllrrrrrrrrr}
				\multicolumn{23}{l}{\begin{tabular}[c]{@{}l@{}}Table \ref{tab:assetallocation} reports the portfolio performance measures for an investor with mean-variance preferences and relative risk-aversion coefficient of five who monthly allocates between\\ equities and risk-free bills using either an historical average (HA) or predictive regression equity risk premium forecast. In panel A, we present the annualized certainty equivalent\\ return gain ($CER_{g}$) in percent for the constructed portfolio based on forecast with either one of the 14 macroeconomic variables or one of the 14 technical indicators. Panel B\\ and C report the portfolio performance measures based on either the PC-ECON, PC-TECH or PC-ALL model for the entire sampling period, business-cycle expansions (exp) \\and recessions (rec). We also compute the portfolio performance measures for the Bullish Index peak periods ($BU^{+}$) and trough periods ($BU^{-}$) as described in the text. The\\ corresponding stable periods are denoted as $stable^{+}$ and $stable^{-}$, respectively. The sampling period spans from 1966:01 to 2019:12.\end{tabular}}                                                                                                              \\ \hline
				\multicolumn{11}{c}{Macroeconomic variables}                                                                                                                                                                                                                                                                                                                                                                                                                                                                                                                                                &  & \multicolumn{11}{c}{Technical indicators}                                                                                                                                                                                                                                                                                                                                                                                                                                                                                                                               \\ \cline{1-11} \cline{13-23} 
				&                            &                      &                                                                             &                                                                            &  &                                                                                    &                                                                                &  &                                                                                    &                                                                                &  &                            &                            &  &                                                                             &                                                                            &  &                                                                                    &                                                                                &  &                                                                                    &                                                                                \\
				\multicolumn{2}{c}{Entire period}                       &                      & \multicolumn{2}{c}{\begin{tabular}[c]{@{}c@{}}Recession \\ dummy\end{tabular}}                                                                           &  & \multicolumn{2}{c}{\begin{tabular}[c]{@{}c@{}}Bullish Index\\  peak\end{tabular}}                                                                                      &  & \multicolumn{2}{c}{\begin{tabular}[c]{@{}c@{}}Bullish Index \\ trough\end{tabular}}                                                                                    &  & \multicolumn{2}{c}{Entire period}                       &  & \multicolumn{2}{c}{\begin{tabular}[c]{@{}c@{}}Recession \\ dummy\end{tabular}}                                                                           &  & \multicolumn{2}{c}{\begin{tabular}[c]{@{}c@{}}Bullish Index \\ peak\end{tabular}}                                                                                      &  & \multicolumn{2}{c}{\begin{tabular}[c]{@{}c@{}}Bullish Index \\ trough\end{tabular}}                                                                                    \\ \cline{1-2} \cline{4-5} \cline{7-8} \cline{10-11} \cline{13-14} \cline{16-17} \cline{19-20} \cline{22-23} 
				\multirow{2}{*}{Predictor} & \multirow{2}{*}{$CER_{g}$} &                      & \multirow{2}{*}{\begin{tabular}[c]{@{}l@{}}$CER_{g}$,\\ $exp$\end{tabular}} & \multirow{2}{*}{\begin{tabular}[c]{@{}l@{}}$CER_{g}$\\ $rec$\end{tabular}} &  & \multirow{2}{*}{\begin{tabular}[c]{@{}l@{}}$CER_{g}$,\\ $stable^{+}$\end{tabular}} & \multirow{2}{*}{\begin{tabular}[c]{@{}l@{}}$CER_{g}$,\\ $BU^{+}$\end{tabular}} &  & \multirow{2}{*}{\begin{tabular}[c]{@{}l@{}}$CER_{g}$,\\ $stable^{-}$\end{tabular}} & \multirow{2}{*}{\begin{tabular}[c]{@{}l@{}}$CER_{g}$,\\ $BU^{-}$\end{tabular}} &  & \multirow{2}{*}{Predictor} & \multirow{2}{*}{$CER_{g}$} &  & \multirow{2}{*}{\begin{tabular}[c]{@{}l@{}}$CER_{g}$,\\ $exp$\end{tabular}} & \multirow{2}{*}{\begin{tabular}[c]{@{}l@{}}$CER_{g}$\\ $rec$\end{tabular}} &  & \multirow{2}{*}{\begin{tabular}[c]{@{}l@{}}$CER_{g}$,\\ $stable^{+}$\end{tabular}} & \multirow{2}{*}{\begin{tabular}[c]{@{}l@{}}$CER_{g}$,\\ $BU^{+}$\end{tabular}} &  & \multirow{2}{*}{\begin{tabular}[c]{@{}l@{}}$CER_{g}$,\\ $stable^{-}$\end{tabular}} & \multirow{2}{*}{\begin{tabular}[c]{@{}l@{}}$CER_{g}$,\\ $BU^{-}$\end{tabular}} \\
				&                            &                      &                                                                             &                                                                            &  &                                                                                    &                                                                                &  &                                                                                    &                                                                                &  &                            &                            &  &                                                                             &                                                                            &  &                                                                                    &                                                                                &  &                                                                                    &                                                                                \\ \hline
				\multicolumn{23}{c}{}                                                                                                                                                                                                                                                                                                                                                                                                                                                                                                                                                                                                                                                                                                                                                                                                                                                                                                                                                                                                                                                                                                                                                    \\
				\multicolumn{23}{c}{Panel A: Bivariate predictive regression forecasts}                                                                                                                                                                                                                                                                                                                                                                                                                                                                                                                                                                                                                                                                                                                                                                                                                                                                                                                                                                                                                                                                                                  \\
				D/P                        & -0.56                      &                      & -1.21                                                                       & 2.94                                                                       &  & -0.99                                                                              & 3.85                                                                           &  & -0.53                                                                              & -0.91                                                                          &  & MA(1,9)                    & 1.28                       &  & -0.62                                                                       & 12.64                                                                      &  & 1.23                                                                               & 1.73                                                                           &  & 1.05                                                                               & 3.43                                                                           \\
				D/Y                        & -0.17                      &                      & -1.40                                                                       & 6.89                                                                       &  & -0.58                                                                              & 4.09                                                                           &  & -0.22                                                                              & 0.24                                                                           &  & MA(1,12)                   & 2.21                       &  & -0.54                                                                       & 18.82                                                                      &  & 2.25                                                                               & 1.84                                                                           &  & 1.77                                                                               & 6.32                                                                           \\
				E/P                        & 0.26                       &                      & -0.12                                                                       & 2.43                                                                       &  & 0.28                                                                               & 0.15                                                                           &  & 0.28                                                                               & 0.08                                                                           &  & MA(2,9)                    & 1.81                       &  & -0.36                                                                       & 14.88                                                                      &  & 1.87                                                                               & 1.23                                                                           &  & 1.50                                                                               & 4.74                                                                           \\
				D/E                        & -0.30                      &                      & -1.21                                                                       & 4.98                                                                       &  & -0.52                                                                              & 1.84                                                                           &  & -0.29                                                                              & -0.44                                                                          &  & MA(2,12)                   & 2.77                       &  & 0.02                                                                        & 19.34                                                                      &  & 2.95                                                                               & 0.99                                                                           &  & 2.36                                                                               & 6.63                                                                           \\
				ERPV                       & -0.82                      &                      & -0.54                                                                       & -2.97                                                                      &  & -1.11                                                                              & 1.99                                                                            &  & -0.32                                                                              & -5.52                                                                          &  & MA(3,9)                    & 1.95                       &  & -0.57                                                                       & 17.09                                                                      &  & 2.02                                                                               & 1.24                                                                           &  & 1.52                                                                               & 5.95                                                                           \\
				B/M                        & -1.20                      & \multicolumn{1}{r}{} & -0.28                                                                       & -6.69                                                                      &  & -1.05                                                                              & -2.64                                                                          &  & -0.91                                                                              & -3.85                                                                          &  & MA(3,12)                   & 1.02                       &  & -0.53                                                                       & 10.39                                                                      &  & 1.18                                                                               & -0.59                                                                          &  & 0.68                                                                               & 4.22                                                                           \\
				NEER                       & 0.18                       &                      & 0.62                                                                        & -2.45                                                                      &  & -0.23                                                                              & 4.27                                                                           &  & -0.08                                                                              & 2.56                                                                           &  & MOM(9)                     & 1.06                       &  & -0.48                                                                       & 10.34                                                                      &  & 1.17                                                                               & -0.05                                                                          &  & 0.70                                                                               & 4.41                                                                           \\
				TBR                        & 1.82                       &                      & 1.15                                                                        & 5.81                                                                       &  & 2.05                                                                               & -0.41                                                                          &  & 1.96                                                                               & 0.63                                                                           &  & MOM(12)                    & 1.02                       &  & -0.46                                                                       & 9.92                                                                       &  & 1.13                                                                               & -0.07                                                                          &  & 0.78                                                                               & 3.20                                                                           \\
				LTY                        & 1.69                       &                      & 0.72                                                                        & 7.40                                                                       &  & 1.95                                                                               & -0.89                                                                          &  & 1.66                                                                               & 1.96                                                                           &  & VOL(1,9)                   & 1.31                       &  & -0.68                                                                       & 13.21                                                                      &  & 1.43                                                                               & 0.06                                                                           &  & 1.03                                                                               & 3.85                                                                           \\
				LTR                        & 1.26                       & \multicolumn{1}{r}{} & -0.11                                                                       & 9.17                                                                       &  & 0.94                                                                               & 4.54                                                                           &  & 0.71                                                                               & 6.33                                                                           &  & VOL(1,12)                  & 2.07                       &  & -0.42                                                                       & 17.02                                                                      &  & 2.35                                                                               & -0.68                                                                          &  & 1.63                                                                               & 6.09                                                                           \\
				TMS                        & 1.63                       &                      & 0.37                                                                        & 8.95                                                                       &  & 1.30                                                                               & 4.84                                                                           &  & 1.64                                                                               & 1.50                                                                           &  & VOL(2,9)                   & 0.94                       &  & -0.39                                                                       & 8.95                                                                       &  & 1.03                                                                               & 0.01                                                                           &  & 0.72                                                                               & 3.01                                                                           \\
				DYS                        & -0.75                      &                      & -0.21                                                                       & -4.01                                                                      &  & -1.00                                                                              & 1.78                                                                           &  & -0.73                                                                              & -0.91                                                                          &  & VOL(2,12)                  & 0.88                       &  & -0.13                                                                       & 6.97                                                                       &  & 1.10                                                                               & -1.30                                                                          &  & 0.79                                                                               & 1.75                                                                           \\
				DRS                        & 0.38                       &                      & 0.73                                                                        & -1.68                                                                      &  & 0.21                                                                               & 2.10                                                                           &  & -0.08                                                                              & 4.63                                                                           &  & VOL(3,9)                   & 0.39                       &  & -0.38                                                                       & 5.06                                                                       &  & 0.54                                                                               & -1.11                                                                          &  & 0.38                                                                               & 0.50                                                                           \\
				INFL                       & 0.31                       &                      & 0.01                                                                        & 2.17                                                                       &  & 0.33                                                                               & 0.18                                                                           &  & 0.24                                                                               & 0.99                                                                           &  & VOL(3,12)                  & 1.70                       &  & -0.29                                                                       & 13.69                                                                      &  & 1.97                                                                               & -0.92                                                                          &  & 1.42                                                                               & 4.35                                                                           \\
				\multicolumn{23}{c}{}                                                                                                                                                                                                                                                                                                                                                                                                                                                                                                                                                                                                                                                                                                                                                                                                                                                                                                                                                                                                                                                                                                                                                    \\
				\multicolumn{23}{c}{Panel B: Principal component predictive regression forecasts}                                                                                                                                                                                                                                                                                                                                                                                                                                                                                                                                                                                                                                                                                                                                                                                                                                                                                                                                                                                                                                                                                        \\
				PC-ECON                    & 1.59                       &                      & -1.17                                                                       & 11.61                                                                      &  & 1.03                                                                               & 7.21                                                                           &  & 1.49                                                                               & 2.53                                                                           &  & PC-TECH                    & 1.94                       &  & -0.44                                                                       & 16.26                                                                      &  & 2.06                                                                               & 0.69                                                                           &  & 1.47                                                                               & 6.30                                                                           \\
				\multicolumn{23}{c}{}                                                                                                                                                                                                                                                                                                                                                                                                                                                                                                                                                                                                                                                                                                                                                                                                                                                                                                                                                                                                                                                                                                                                                    \\
				\multicolumn{23}{c}{Panel C: Principal component predictive regression forecasts, all predictors taken together}                                                                                                                                                                                                                                                                                                                                                                                                                                                                                                                                                                                                                                                                                                                                                                                                                                                                                                                                                                                                                                                                    \\
				PC-ALL                     & 3.94                       &                      & 0.24                                                                        & 26.32                                                                      &  & 3.59                                                                               & 7.52                                                                           &  & 3.31                                                                               & 9.81                                                                           &  &                            &                            &  &                                                                             &                                                                            &  &                                                                                    &                                                                                &  &                                                                                    &                                                                                \\ \hline
			\end{tabular}%
		}
	\end{table}
\end{landscape}
%\hspace*{-1.5cm}\resizebox{26cm} \thispagestyle{empty}
\newpage
% \usepackage{multirow}
% \usepackage{graphicx}
% \usepackage{multirow}
% \usepackage{graphicx}
% \usepackage{multirow}
% \usepackage{graphicx}
% \usepackage{lscape}
% \usepackage{multirow}
% \usepackage{graphicx}
% \usepackage{lscape}
\begin{landscape}
	\begin{table}[th]
		\centering
		\caption{Certainty equivalent return gain ($CER_{g}$) over  different holding period}
		\label{tab:holding periods}
		\thispagestyle{empty}
	\hspace*{-1.5cm}\resizebox{26cm}{!}{%
			\begin{tabular}{llrrrrrrrrrrrrr}
				\multicolumn{15}{l}{\begin{tabular}[c]{@{}l@{}}Table \ref{tab:holding periods} reports the net-of-transactions-costs annualized certainty equivalent return gain ($CER_{g}$) in percent for different holding periods ($G_{t}$) in months subsequently after the\\ Bullish Index reaches its peak (trough) as described in the text, we denote the corresponding stable periods as $stable^{+}$ ($stable^{-}$). The portfolio is constructed based on either one\\ of the PC-ECON, PC-TECH, or PC-ALL model. The sampling period spans from 1966:01 to 2019:12.\end{tabular}}                                                                                                                                                                                                                                                                                                                                                                                                                                                                                                                                           \\ \hline
				&  & \multicolumn{6}{c}{Bullish Index peak}                                                                                                                                                                                                                                                                                                                                                                                                                                                               &  & \multicolumn{6}{c}{Bullish Index trough}                                                                                                                                                                                                                                                                                                                                                                                                                                                             \\ \cline{3-8} \cline{10-15} 
				\multirow{3}{*}{\begin{tabular}[c]{@{}l@{}}Holding \\ period (month)\end{tabular}} &  & \multirow{3}{*}{\begin{tabular}[c]{@{}l@{}}PC-ECON,\\ $stable^{+}$\end{tabular}} & \multirow{3}{*}{\begin{tabular}[c]{@{}l@{}}PC-ECON,\\ $BU^{+}$\end{tabular}} & \multirow{3}{*}{\begin{tabular}[c]{@{}l@{}}PC-TECH,\\ $stable^{+}$\end{tabular}} & \multirow{3}{*}{\begin{tabular}[c]{@{}l@{}}PC-TECH,\\ $BU^{+}$\end{tabular}} & \multirow{3}{*}{\begin{tabular}[c]{@{}l@{}}PC-ALL,\\ $stable^{+}$\end{tabular}} & \multirow{3}{*}{\begin{tabular}[c]{@{}l@{}}PC-ALL,\\ $BU^{+}$\end{tabular}} &  & \multirow{3}{*}{\begin{tabular}[c]{@{}l@{}}PC-ECON,\\ $stable^{-}$\end{tabular}} & \multirow{3}{*}{\begin{tabular}[c]{@{}l@{}}PC-ECON,\\ $BU^{-}$\end{tabular}} & \multirow{3}{*}{\begin{tabular}[c]{@{}l@{}}PC-TECH,\\ $stable^{-}$\end{tabular}} & \multirow{3}{*}{\begin{tabular}[c]{@{}l@{}}PC-TECH,\\ $BU^{-}$\end{tabular}} & \multirow{3}{*}{\begin{tabular}[c]{@{}l@{}}PC-ALL,\\ $stable^{-}$\end{tabular}} & \multirow{3}{*}{\begin{tabular}[c]{@{}l@{}}PC-ALL,\\ $BU^{-}$\end{tabular}} \\
				&  &                                                                                  &                                                                              &                                                                                  &                                                                              &                                                                                 &                                                                             &  &                                                                                  &                                                                              &                                                                                  &                                                                              &                                                                                 &                                                                             \\
				&  &                                                                                  &                                                                              &                                                                                  &                                                                              &                                                                                 &                                                                             &  &                                                                                  &                                                                              &                                                                                  &                                                                              &                                                                                 &                                                                             \\ \hline
				$G_{1} \rightarrow G_{3}$                                                          &  & 0.15                                                                             & 6.97                                                                         & 1.65                                                                             & 0.14                                                                         & 2.64                                                                            & 7.08                                                                        &  & 0.61                                                                             & 1.86                                                                         & 1.07                                                                             & 5.55                                                                         & 2.41                                                                            & 9.02                                                                        \\
				$G_{4} \rightarrow G_{6}$                                                          &  & -0.75                                                                            & 15.56                                                                        & 1.51                                                                             & 1.69                                                                         & 1.69                                                                            & 16.43                                                                       &  & 0.37                                                                             & 3.80                                                                         & 0.58                                                                             & 9.85                                                                         & 1.50                                                                            & 16.85                                                                       \\
				$G_{7} \rightarrow G_{9}$                                                          &  & 0.67                                                                             & 1.12                                                                         & 1.72                                                                             & -0.33                                                                        & 2.96                                                                            & 3.69                                                                        &  & 1.41                                                                             & -5.70                                                                        & 1.30                                                                             & 3.42                                                                         & 3.74                                                                            & -2.90                                                                       \\
				$G_{10} \rightarrow G_{12}$                                                        &  & 0.62                                                                             & 1.49                                                                         & 1.18                                                                             & 5.36                                                                         & 2.62                                                                            & 7.05                                                                        &  & 0.60                                                                             & 1.58                                                                         & 1.56                                                                             & 1.08                                                                         & 2.95                                                                            & 3.80                                                                        \\ \hline
				\multicolumn{15}{l}{}                                                                                                                                                                                                                                                                                                                                                                                                                                                                                                                                                                                                                                                                                                                                                                                                                                                                                                                                                                                                                                                                           
			\end{tabular}%
		}
	\end{table}
\end{landscape}
\newpage

\begin{table}[th]
	\centering
	\caption{Robustness check  (excluding recession dummy)}
	\label{tab:Robustness exclude recession}
	\resizebox{\textwidth}{!}{%
		\begin{tabular}{lrrrrrrrrrlrrrrrrrr}
			\multicolumn{19}{l}{\begin{tabular}[c]{@{}l@{}}Table \ref{tab:Robustness exclude recession} reports the portfolio performance for a mean-variance investor with a relative risk-aversion coefficient of five who \\ monthly allocates between equities and risk-free bills using either a historical average (HA) or predictive regression equity risk \\ premium forecast. In Panel A, we present the net-of-transactions-costs annualized certainty equivalent return gain ($CER_{g}$) in\\ percent for the constructed portfolio based on either one of the  14 macroeconomic variables or one of the 14 technical indicators\\ for the Bullish Index peak periods ($BU^{+}$) and trough periods ($BU^{-}$) as described in the text. The corresponding stable periods\\ are denoted as $stable^{+}$ and $stable^{-}$, respectively. Panel B and C report the net-of-transactions-costs portfolio performance\\ measures based on either the PC-ECON, PC-TECH or PC-ALL model. The sampling period spans from 1966:01 to 2019:12.\end{tabular}} \\ \hline
			\multicolumn{9}{c}{\multirow{2}{*}{Macroeconomic variables}}                                                                                                                                                                                                                                                                                                                                                                                                                                                                                                                                                            &                       & \multicolumn{9}{c}{\multirow{2}{*}{Technical indicators}}                                                                                                                                                                                                                                                                                                                                                                                                                                                                                                                                                           \\
			\multicolumn{9}{c}{}                                                                                                                                                                                                                                                                                                                                                                                                                                                                                                                                                                                                    &                       & \multicolumn{9}{c}{}                                                                                                                                                                                                                                                                                                                                                                                                                                                                                                                                                                                                \\ \cline{1-9} \cline{11-19} 
			\multicolumn{1}{c}{}                             & \multicolumn{1}{c}{}                       & \multicolumn{3}{c}{Bullish Index peak}                                                                                                                                                                                               &                                           & \multicolumn{3}{c}{Bullish Index trough}                                                                                                                                                                                           &                       & \multicolumn{1}{c}{}                            & \multicolumn{1}{c}{}                      & \multicolumn{3}{c}{Bullish Index peak}                                                                                                                                                                                             & \multicolumn{1}{c}{}                      & \multicolumn{3}{c}{Bullish Index trough}                                                                                                                                                                                           \\ \cline{3-5} \cline{7-9} \cline{13-15} \cline{17-19} 
			\multirow{2}{*}{Predictor}                       &                                            & \multirow{2}{*}{\begin{tabular}[c]{@{}l@{}}$CER_{g}$,\\ $stable^{+}$\end{tabular}}                       &                        & \multirow{2}{*}{\begin{tabular}[c]{@{}l@{}}$CER_{g}$,\\ $BU^{+}$\end{tabular}}                      &                                           & \multirow{2}{*}{\begin{tabular}[c]{@{}l@{}}$CER_{g}$,\\ $stable^{-}$\end{tabular}}                      &                       & \multirow{2}{*}{\begin{tabular}[c]{@{}l@{}}$CER_{g}$,\\ $BU^{-}$\end{tabular}}                      &                       & \multirow{2}{*}{Predictor}                      &                                           & \multirow{2}{*}{\begin{tabular}[c]{@{}l@{}}$CER_{g}$,\\ $stable^{+}$\end{tabular}}                      &                       & \multirow{2}{*}{\begin{tabular}[c]{@{}l@{}}$CER_{g}$,\\ $BU^{+}$\end{tabular}}                      &                                           & \multirow{2}{*}{\begin{tabular}[c]{@{}l@{}}$CER_{g}$,\\ $stable^{-}$\end{tabular}}                      &                       & \multirow{2}{*}{\begin{tabular}[c]{@{}l@{}}$CER_{g}$,\\ $BU^{-}$\end{tabular}}                      \\
			&                                            &                                                                                                         &                        &                                                                                                    &                                           &                                                                                                        &                       &                                                                                                    &                       &                                                 &                                           &                                                                                                        &                       &                                                                                                    &                                           &                                                                                                        &                       &                                                                                                    \\ \hline
			\multicolumn{19}{c}{}                                                                                                                                                                                                                                                                                                                                                                                                                                                                                                                                                                                                                                                                                                                                                                                                                                                                                                                                                                                                                                                                                                                                                                                                                                                 \\
			\multicolumn{19}{c}{Panel A: Bivariate predictive regression forecasts}                                                                                                                                                                                                                                                                                                                                                                                                                                                                                                                                                                                                                                                                                                                                                                                                                                                                                                                                                                                                                                                                                                                                                                                               \\
			D/P                                               &                                            & -0.81                                                                                                   &                        & 1.95                                                                                               &                                           & -0.78                                                                                                  &                       & 0.95                                                                                               &                       & MA(1,9)                                         &                                           & 0.67                                                                                                   &                       & 1.94                                                                                               &                                           & 0.78                                                                                                   &                       & 0.67                                                                                               \\
			D/Y                                               &                                            & -0.55                                                                                                   &                        & 2.57                                                                                               &                                           & -0.46                                                                                                  &                       & 0.71                                                                                               &                       & MA(1,12)                                        &                                           & 1.69                                                                                                   &                       & 1.84                                                                                               &                                           & 1.70                                                                                                   &                       & 1.99                                                                                               \\
			E/P                                               &                                            & 0.16                                                                                                    &                        & 0.89                                                                                               &                                           & 0.10                                                                                                   &                       & 1.40                                                                                               &                       & MA(2,9)                                         &                                           & 1.33                                                                                                   &                       & 1.35                                                                                               &                                           & 1.35                                                                                                   &                       & 1.47                                                                                               \\
			D/E                                               &                                            & -0.58                                                                                                   &                        & 1.76                                                                                               &                                           & -0.35                                                                                                  &                       & -1.66                                                                                              &                       & MA(2,12)                                        &                                           & 2.47                                                                                                   &                       & 0.83                                                                                               &                                           & 2.39                                                                                                   &                       & 2.41                                                                                               \\
			ERPV                                             &                                            & -1.42                                                                                                   &                        & 2.03                                                                                               &                                           & -1.11                                                                                                  &                       & -2.21                                                                                              &                       & MA(3,9)                                         &                                           & 1.51                                                                                                   &                       & 0.58                                                                                               &                                           & 1.37                                                                                                   &                       & 2.49                                                                                               \\
			B/M                                               &                                            & -1.33                                                                                                   &                        & -2.01                                                                                              & \multicolumn{1}{r}{}                      & -1.57                                                                                                  &                       & 1.16                                                                                               &                       & MA(3,12)                                        &                                           & 0.77                                                                                                   &                       & 0.62                                                                                               &                                           & 0.67                                                                                                   &                       & 1.98                                                                                               \\
			NEER                                             &                                            & -0.48                                                                                                   &                        & 5.60                                                                                               &                                           & -0.34                                                                                                  &                       & 2.70                                                                                               &                       & MOM(9)                                          &                                           & 0.80                                                                                                   &                       & 0.74                                                                                               &                                           & 0.68                                                                                                   &                       & 2.11                                                                                               \\
			TBR                                              &                                            & 1.77                                                                                                    &                        & 2.27                                                                                               &                                           & 2.07                                                                                                   &                       & -1.43                                                                                              &                       & MOM(12)                                         &                                           & 0.88                                                                                                   &                       & -0.18                                                                                              &                                           & 0.82                                                                                                   &                       & 0.72                                                                                               \\
			LTY                                              &                                            & 1.74                                                                                                    &                        & 1.10                                                                                               &                                           & 1.83                                                                                                   &                       & 0.26                                                                                               &                       & VOL(1,9)                                        &                                           & 0.69                                                                                                   &                       & 0.87                                                                                               &                                           & 0.62                                                                                                   &                       & 1.44                                                                                               \\
			LTR                                              &                                            & -1.74                                                                                                   &                        & 0.46                                                                                               & \multicolumn{1}{r}{}                      & -1.39                                                                                                  &                       & -3.11                                                                                              &                       & VOL(1,12)                                       &                                           & 1.56                                                                                                   &                       & 1.02                                                                                               &                                           & 1.42                                                                                                   &                       & 2.44                                                                                               \\
			TMS                                              &                                            & 1.10                                                                                                    &                        & 2.78                                                                                               &                                           & 1.39                                                                                                   &                       & -1.37                                                                                              &                       & VOL(2,9)                                        &                                           & 0.62                                                                                                   &                       & 0.84                                                                                               &                                           & 0.52                                                                                                   &                       & 1.73                                                                                               \\
			DYS                                              &                                            & -1.09                                                                                                   &                        & 1.35                                                                                               &                                           & -0.95                                                                                                  &                       & -0.86                                                                                              &                       & VOL(2,12)                                       &                                           & 0.66                                                                                                   &                       & 0.47                                                                                               &                                           & 0.61                                                                                                   &                       & 0.96                                                                                               \\
			DRS                                              &                                            & -0.91                                                                                                   &                        & 0.81                                                                                               &                                           & -0.97                                                                                                  &                       & 1.51                                                                                               &                       & VOL(3,9)                                        &                                           & 0.18                                                                                                   &                       & 0.36                                                                                               &                                           & 0.17                                                                                                   &                       & 0.36                                                                                               \\
			INFL                                             &                                            & -0.59                                                                                                   &                        & 0.67                                                                                               &                                           & -0.33                                                                                                  &                       & -2.49                                                                                              &                       & VOL(3,12)                                       &                                           & 1.46                                                                                                   &                       & 0.69                                                                                               &                                           & 1.43                                                                                                   &                       & 0.84                                                                                               \\
			\multicolumn{19}{c}{}                                                                                                                                                                                                                                                                                                                                                                                                                                                                                                                                                                                                                                                                                                                                                                                                                                                                                                                                                                                                                                                                                                                                                                                                                                                 \\
			\multicolumn{19}{c}{Panel B: Principal component predictive regression forecasts}                                                                                                                                                                                                                                                                                                                                                                                                                                                                                                                                                                                                                                                                                                                                                                                                                                                                                                                                                                                                                                                                                                                                                                                     \\
			PC-ECON                                &                                            & 0.39                                                                                                    &                        & 6.44                                                                                               &                                           & 0.90                                                                                                   &                       & -1.01                                                                                              &                       & PC-TECH                                &                                           & 1.56                                                                                                   &                       & 1.45                                                                                               &                                           & 1.47                                                                                                   &                       & 2.50                                                                                               \\
			\multicolumn{19}{c}{}                                                                                                                                                                                                                                                                                                                                                                                                                                                                                                                                                                                                                                                                                                                                                                                                                                                                                                                                                                                                                                                                                                                                                                                                                                                 \\
			\multicolumn{19}{c}{Panel C: Principal component predictive regression forecasts, all predictors taken together}                                                                                                                                                                                                                                                                                                                                                                                                                                                                                                                                                                                                                                                                                                                                                                                                                                                                                                                                                                                                                                                                                                                                                                 \\
			PC-ALL                                  &                                            & 2.80                                                                                                    &                        & 6.80                                                                                               &                                           & 3.16                                                                                                   &                       & 1.92                                                                                               &                       &                                                 &                                           &                                                                                                        &                       &                                                                                                    &                                           &                                                                                                        &                       &                                                                                                    \\ \hline
		\end{tabular}%
	}
\end{table}

% \usepackage{graphicx}

\newpage

\newpage
\FloatBarrier

\newpage
\FloatBarrier


\clearpage


\begin{thebibliography}{99}

\bibitem{ivanova1999low} K. Ivanova, M. Ausloos, Low-order variability diagrams for short-range correlation evidence in financial data: BGL-USD exchange rate, Dow Jones industrial average, gold ounce price, Physica A: Statistical
Mechanics and its Applications 265 (1-2) (1999) 279-291, doi:10.1016/S0378-4371(98)00562-7.

\bibitem{ausloos2000statistical} M. Ausloos, Statistical physics in foreign exchange currency and stock markets, Physica A: Statistical Mechanics and its Applications 285 (1) (2000) 48-65, doi:10.1016/S0378-4371(00)00271-5.

\bibitem{ivanova2002eur} M. Ausloos, K. Ivanova, Are EUR and GBP diffferent words for the same currency?, The
European Physical Journal B-Condensed Matter and Complex Systems 27 (2) (2002) 239-247,
doi:10.1140/epjb/e20020155.

\bibitem{calvet2002multifractality} L. Calvet, A. Fisher, Multifractality in asset returns: theory and evidence, Review of Economics and Statistics 84 (3) (2002) 381-406, doi:10.1162/003465302320259420.

\bibitem{BuonocoreBrandiMantengaDiMatteo2020} R. J. Buonocore, G. Brandi, R. N. Mantegna, T. Di Matteo, On the interplay between multiscaling and
stock dependence, Quantitative Finance 20 (1) (2020) 133-145, doi:10.1080/14697688.2019.1645345.

\bibitem{KukackaKristoufek2020} J. Kukacka, L. Kristoufek, Do 'complex' financial models really lead to complex dynamics? Agent based models and multifractality, Journal of Economic Dynamics and Control 113 (2020) 103855,
 doi:10.1016/j.jedc.2020.103855.

\bibitem{BrandiDiMatteo2021} G. Brandi, T. Di Matteo, On the statistics of scaling exponents and the Multiscaling Value at Risk, European Journal of Finance 28 (13-15) (2022) 1361-1382, doi:10.1080/1351847X.2021.1908391.

\bibitem{chan2021economic}
J. Chan Phooi M'ng, H. Y. Jer, Do economic statistics contain information to predict stock indexes futures prices and returns? Evidence from Asian equity futures markets, Review of Quantitative Finance  and Accounting 57 (3) (2021) 1033-1060, doi:10.1007/s11156-021-00969-2.

\bibitem{hurst1951long}. H. E. Hurst, Long-term storage capacity of reservoirs, Transactions of the American Society of Civil
Engineers 116 (1951) 770-799.

\bibitem{jiang2019multifractal} Z.-Q. Jiang, W.-J. Xie, W.-X. Zhou, D. Sornette, Multifractal analysis of financial markets: A Review, Reports on Progress in Physics 82 (2019) 125901, doi:10.1088/1361-6633/ab42fb.

\bibitem{wu2017fractal} B. Wu, T. Duan, The fractal feature and price trend in the gold future market at the Shanghai
Futures Exchange (SFE), Physica A: Statistical Mechanics and its Applications 474 (2017) 99-106,
doi:10.1016/j.physa.2016.12.048.

\bibitem{vandewalle1997coherent}
  N. Vandewalle, M. Ausloos, Coherent and random sequences in financial fluctuations, Physica A: Statistical
Mechanics and its Applications 246 (3) (1997) 454-459, doi:10.1016/S0378-4371(97)00366-X.

\bibitem{ausloos2004classical}  M. Ausloos, K. Ivanova, Classical technical analysis of Latin American market indices: correlations in Latin American currencies (ARS, CLP, MXP) exchange rates with respect to DEM, GBP, JPY and
USD, Brazilian Journal of Physics 34 (2A) (2004) 504-511, doi:10.1590/S0103-97332004000300029.

\bibitem{cristescu2012parameter}
C. P. Cristescu, C. Stan, E. I. Scarlat, T. Minea, C. M. Cristescu, Parameter motivated mutual correlation
analysis: Application to the study of currency exchange rates based on intermittency parameter
  and Hurst exponent, Physica A: Statistical Mechanics and its Applications 391 (8) (2012) 2623-2635,
doi:10.1016/j.physa.2011.12.006.

\bibitem{ausloos2002multifractal} M. Ausloos, K. Ivanova, Multifractal nature of stock exchange prices, Computer Physics Communications 147 (1-2) (2002) 582-585.

\bibitem{alvarez2008time} J. Alvarez-Ramirez, J. Alvarez, E. Rodriguez, G. Fernandez-Anaya, Time-varying Hurst exponent for US stock markets, Physica A: Statistical Mechanics and its Applications 387 (24) (2008) 6159-6169,  https://doi.org/10.1016/j.physa.2008.06.056

%doi:10.1016/j.jemp n.2019.09.001.

\bibitem{wang2009analysis}
Y. Wang, L. Liu, R. Gu, Analysis of efficiency for Shenzhen stock market based on multifractal detrended fluctuation analysis, International Review of Financial Analysis 18 (5) (2009) 271-276, doi:10.1016/j.irfa.2009.09.005.

\bibitem{zhu2018multifractal}  H. Zhu,  W. Zhang, Multifractal property of Chinese stock market in the CSI 800 index based on MF-DFA approach, Physica A: Statistical Mechanics and its Applications 490 (2018) 497-503,
doi:10.1016/j.physa.2017.08.060.

\bibitem{yang2016multifractal} L. Yang, Y. Zhu, Y. Wang, Y. Wang, Multifractal detrended cross-correlations between crude oil market and Chinese ten sector stock markets, Physica A: Statistical Mechanics and its Applications 462 (2016) 255-265, doi:10.1016/j.physa.2016.06.040.

\bibitem{watorek2019multifractal}
M. Watorek, S. Drozdz, P. Oswiecimka, M. Stanuszek, Multifractal cross-correlations between the world oil and other financial markets in 2012-2017, Energy Economics 81 (2019) 874-885, doi:10.1016/j.eneco.2019.05.015.

\bibitem{yao2020multifractal}  C.-Z. Yao, C. Liu, W.-J. Ju, Multifractal analysis of the WTI crude oil market, US stock
 market and EPU, Physica A: Statistical Mechanics and its Applications 550 (2020) 124096, doi:10.1016/j.physa.2019.124096.

\bibitem{takaishi2018statistical} T. Takaishi, Statistical properties and multifractality of Bitcoin, Physica A: Statistical Mechanics and its Applications 506 (2018) 507-519, doi:10.1016/j.physa.2018.04.046.

\bibitem{zhangstylised}  Y. Zhang, S. Chan, J. Chu, S. Nadarajah, Stylised facts for high frequency cryptocurrency data, Physica A: Statistical Mechanics and its Applications 513 (2019) 598-612, doi:10.1016/j.physa.2018.09.042.

\bibitem{vandewalle1998crossing} N. Vandewalle, M. Ausloos, Crossing of two mobile averages: A method for measuring the roughness exponent, Physical Review E 58 (5) (1998) 6832, doi:10.1103/PhysRevE.58.6832.

\bibitem{di2005long} T. Di Matteo, T. Aste, M. M. Dacorogna, Long-term memories of developed and emerging markets:
Using the scaling analysis to characterize their stage of development, Journal of Banking \& Finance, 29 (4) (2005) 827-851, doi:10.1016/j.jbank-n.2004.08.004.

\bibitem{carbone2009detrending} A. Carbone, Detrending moving average algorithm: a brief review, in: 2009 IEEE Toronto International Conference Science and Technology for Humanity (TIC-STH), IEEE, 2009, pp. 691-696, doi:10.1109/TIC-STH.2009.5444412.

\bibitem{gu2010detrending} G.-F. Gu, W.-X. Zhou, Detrending moving average algorithm for multifractals, Physical Review E 82 (1),  (2010) 011136, doi:10.1103/PhysRevE.82.011136.

\bibitem{chen2016finite}  F. Chen, K. Tian, X. Ding, Y. Miao, C. Lu, Finite-size effect and the components of multifractality
in transport economics volatility based on multifractal detrending moving average method, Physica A:
Statistical Mechanics and its Applications 462 (2016) 1058-1066, doi:10.1016/j.physa.2016.06.101.

\bibitem{auer2016performance} B. R. Auer, On the performance of simple trading rules derived from the fractal dynamics of gold and silver price fluctuations, Finance Research Letters 16 (2016) 255-267, doi:10.1016/j.frl.2015.12.009

\bibitem{eom2008hurst} C. Eom, S. Choi, G. Oh, W.-S. Jung, Hurst exponent and prediction based on weak-form efficient
market hypothesis of stock markets, Physica A: Statistical Mechanics and its Applications 387 (18) (2008) 4630-4636, doi:10.1016/j.physa.2008.03.035.

\bibitem{grech2004can} D. Grech, Z. Mazur, Can one make any crash prediction in finance using the local Hurst exponent idea?, Physica A: Statistical Mechanics and its Applications 336 (1-2) (2004) 133-145, doi:10.1016/j.physa.2004.01.018.

\bibitem{grech2008local} D. Grech, G. Pamuła, The local Hurst exponent of the financial time series in the vicinity of crashes
on the Polish stock exchange market, Physica A: Statistical Mechanics and its Applications 387 (16-17) (2008) 4299-4308, doi:10.1016/j.physa.2017.06.032.

\bibitem{varela2015long} M. P.   B. Varela, F. Biney, I. Florescu, Long correlations and fractional difference analysis applied to the study of memory effects in high-frequency (tick) data, Quantitative Finance 15 (8) (2015) 1365-1374, doi:10.1080/14697688.2015.1032547.
 
\bibitem{welch2008comprehensive}  I. Welch, A. Goyal, A comprehensive look at the empirical performance of equity premium prediction, Review of Financial Studies 21 (4) (2008) 1455-1508, doi:10.1093/rfs/hhm014.
  
\bibitem{campbell2008predicting} J. Y. Campbell, S. B. Thompson, Predicting excess stock returns out of sample: Can anything beat the historical average?, Review of Financial Studies 21 (4) (2008) 1509-1531, doi:10.1093/rfs/hhm055.
  
\bibitem{peters1994fractal} E. E. Peters, Fractal market analysis: applying chaos theory to investment and economics, John Wiley \& Sons, 1994.

\bibitem{kim2010portfolio} S. Kim, F. In, Portfolio allocation and the investment horizon: a multiscaling approach, Quantitative
595 Finance 10 (4) (2010) 443-453, doi:10.1080/14697680902960226.

\bibitem{domino2011use} K. Domino, The use of the Hurst exponent to predict changes in trends on the Warsaw
Stock Exchange, Physica A: Statistical Mechanics and its Applications 390 (1) (2011) 98-109,
doi:10.1016/j.physa.2010.04.015.

\bibitem{batten2013structure} J. A. Batten, C. Ciner, B. M. Lucey, P. G. Szilagyi, The structure of gold and silver spread returns,  Quantitative Finance 13 (4) (2013) 561-570, doi:10.1080/14697688.2012.708777.

\bibitem{ossadnik1994correlation} S. M. Ossadnik, S. V. Buldyrev, A. L. Goldberger, S. Havlin, R. N. Mantegna, C. K. Peng, M. Simons, H. E. Stanley, Correlation approach to identify coding regions in DNA sequences, Biophysical Journal 67 (1) (1994) 64-70, doi:10.1016/S0006-3495(94)80455-2.

\bibitem{peng1994mosaic} C.-K. Peng, S. V. Buldyrev, S. Havlin, M. Simons, H. E. Stanley, A. L. Goldberger, Mosaic organization of DNA nucleotides, Physical Review E 49 (2) (1994) 1685, doi:10.1103/PhysRevE.49.1685.

\bibitem{alessio2002second} E. Alessio, A. Carbone, G. Castelli, V. Frappietro, Second-order moving average and scaling of stochastic time series, The European Physical Journal B-Condensed Matter and Complex Systems 27 (2) (2002) 197-200, doi:10.1140/epjb/e20020150.

\bibitem{mandelbrot1971analysis} B. Mandelbrot, Analysis of long-run dependence in economics: The R/S technique., Econometrica 39 610 (1971) 68-69.

\bibitem{mantegna1995scaling} R. N. Mantegna, H. E. Stanley, Scaling behaviour in the dynamics of an economic index, Nature 376 (1995) 46-49, doi:10.1038/376046a0.

\bibitem{amaral1997scaling}  L. A. N. Amaral, S. V. Buldyrev, S. Havlin, P. Maass, M. A. Salinger, H. E. Stanley, M. H. Stanley,
Scaling behavior in economics: The problem of quantifying company growth, Physica A: Statistical  Mechanics and its Applications 244 (1-4) (1997) 1-24, doi:10.1016/S0378-4371(97)00301-4.

\bibitem{vandewalle1998multi} N. Vandewalle, M. Ausloos, Multi-affine analysis of typical currency exchange rates, The European Physical Journal B-Condensed Matter and Complex Systems 4 (2) (1998) 257-261, doi:10.1007/s100510050376.

\bibitem{ausloos2002financial} M. Ausloos, Financial time series and statistical mechanics, in: Computational Statistical Physics,  Springer, 2002, pp. 153-168, doi:10.1007/978-3-662-04804-710.

\bibitem{cerqueti2003microeconomic} R. Cerqueti, G. Rotundo, Microeconomic modeling of financial time series with long term memory, in: 2003 IEEE International Conference on Computational Intelligence for Financial Engineering, 2003.
Proceedings., IEEE, 2003, pp. 191-198, doi:10.1109/CIFER.2003.1196260.

\bibitem{baciu2014ranking}  O. A. Baciu, Ranking capital markets efficiency: The case of twenty European stock markets, Journal of Applied Quantitative Methods 9 (3) (2014) 24-33.

\bibitem{gunduz2022entropic} G. Gündüz, M. Kuzucuoğlu, Y. Gündüz, Entropic characterization of Gross Domestic Product per capita (GDP) values of countries, Physica A: Statistical Mechanics and its Applications 603 (2022) 127831, doi:10.1016/j.physa.2022.127831.

\bibitem{tilfani2022heterogeneity} O. Tilfani, L. Kristoufek, P. Ferreira, M. Y. El Boukfaoui, Heterogeneity in economic relationships:  Scale dependence through the multivariate fractal regression, Physica A: Statistical Mechanics and its
Applications 588 (2022) 126530, doi:10.1007/s100510050376.

\bibitem{cajueiro2004hurst} D. O. Cajueiro, B. M. Tabak, The Hurst exponent over time: testing the assertion that emerging
markets are becoming more efficient, Physica A: Statistical Mechanics and its Applications 336 (3-4) (2004) 521-537, doi:10.1016/j.physa.2003.12.031.

\bibitem{jafari2007does}  G. R. Jafari, A. Bahraminasab, P. Norouzzadeh, Why does the Standard GARCH (1, 1) model work well?, International Journal of Modern Physics C 18 (07) (2007) 1223-1230, doi:10.1142/S0129183107011261.

\bibitem{xu2005quantifying}
 L. Xu, P. C. Ivanov, K. Hu, Z. Chen, A. Carbone, H. E. Stanley, Quantifying signals with power-law correlations: A comparative study of detrended fluctuation analysis and detrended moving average techniques, Physical Review E 71 (5) (2005) 051101, doi:10.1103/PhysRevE.71.051101.
 
\bibitem{ausloos2003strategy}  M. Ausloos, P. Bronlet, Strategy for investments from Zipf law(s), Physica A: Statistical Mechanics and its Applications 324 (1) (2003) 30-37, doi:10.1016/S0378-4371(02)01845-9.

\bibitem{horta2014impact} P. Horta, S. Lagoa, L. Martins, The impact of the 2008 and 2010 financial crises on the Hurst exponents of international stock markets: Implications for efficiency and contagion, International Review of  Financial Analysis 35 (2014) 140-153, doi:10.1016/j.irfa.2014.08.002.

\bibitem{barunik2010hurst} J. Barunik, L. Kristoufek, On Hurst exponent estimation under heavy-tailed distributions, Physica A: Statistical Mechanics and its Applications 389 (18) (2010) 3844-3855, doi:10.1016/j.physa.2010.05.025.

\bibitem{ren2019balanced} Y. Ren, Y. Tu, Y. Yi, Balanced predictive regressions, Journal of Empirical Finance 54 (2019) 118-142, https://doi.org/10.1016/j.jempfin.2019.09.001

\bibitem{wang2019oil} Y. Wang, Z. Pan, L. Liu, C. Wu, Oil price increases and the predictability of equity premium, Journal
of Banking \& Finance 102 (2019) 43-58, https://doi.org/10.1016/j.jbankfin.2019.03.009

\bibitem{zhang2019forecasting} Y. Zhang, F. Ma, Y. Wang, Forecasting crude oil prices with a large set of predictors: Can LASSO select powerful predictors?, Journal of Empirical Finance 54 (2019) 97-117,https://doi.org/10.1016/j.jempfin.2019.08.007
  
\bibitem{mele2007asymmetric} A. Mele, Asymmetric stock market volatility and the cyclical behavior of expected returns, Journal of  Financial Economics 86 (2) (2007) 446-478, https://doi.org/10.1016/j.jfineco.2006.10.002
 
\bibitem{neely2014forecasting}  C. J. Neely, D. E. Rapach, J. Tu, G. Zhou, Forecasting the equity risk premium: the role of technical indicators, Management Science 60 (7) (2014) 1772-1791, doi:10.1287/mnsc.2013.1838.
  
\bibitem{sun2021can}  M. Sun, P. Glabadanidis, Can technical indicators predict the Chinese equity risk premium?, International Review of Finance 22 (1) (2022) 114-142, doi:10.1111/ir.12344.
   
\bibitem{clark2007approximately} T. E. Clark, K. D. West, Approximately normal tests for equal predictive accuracy in nested models, Journal of Econometrics 138 (1) (2007) 291-311, doi:10.1093/rfs/hhm055.
    
\bibitem{ferreira2011forecasting} M. A. Ferreira, P. Santa-Clara, Forecasting stock market returns: The sum of the parts is more than the whole, Journal of Financial Economics 100 (3) (2011) 514-537, doi:10.1016/j.jfineco.2011.02.003.
    
\bibitem{jovanovic2018comparison} F. Jovanovic, A comparison between qualitative and quantitative histories: the example of
  the efficient market hypothesis, Journal of Economic Methodology 25 (4) (2018) 291-310, doi:10.1080/1350178X.2018.1529135.

\bibitem{ramos2017introducing} J. P. Ramos-Requena, J. Trinidad-Segovia, M. Sánchez-Granero, Introducing Hurst exponent
in pair trading, Physica A: Statistical Mechanics and its Applications 488 (2017) 39-45, doi:10.1016/j.physa.2017.06.032.

\bibitem{garcia2019different} M. N. López-García, J. P. Ramos-Requena, Different methodologies and uses of the Hurst exponent in econophysics, Estudios de Economia Aplicada 37 (2) (2019) 96-108, doi:10.25115/eea.v37i2.2603.

\bibitem{newey1986simple} W. K. Newey, K. D. West, A simple, positive semi-definite, heteroskedasticity and autocorrelation consistent covariance matrix, Econometrica 55(3) (1987) 703-708, https://www.nber.org/papers/t0055.pdf.
    
\end{thebibliography}
\end{document}